\documentclass[final,1p,times]{elsarticle}

\usepackage{lineno,hyperref}
\usepackage{tabularx}
\usepackage{fancyvrb}
\usepackage{textgreek}
\usepackage{amsmath}
\usepackage{booktabs}
\usepackage{xcolor}
\usepackage{comment}

\modulolinenumbers[5]
\journal{CPC}

\begin{document}

\newcommand{\be}{\begin{equation}}
\newcommand{\ee}{\end{equation}}
\newcommand{\bea}{\begin{eqnarray}}
\newcommand{\eea}{\end{eqnarray}}
\newcommand{\ep}{\varepsilon}
\newcommand{\mi}{\mathcal{T}}

\newcolumntype{C}[1]{>{\centering\arraybackslash}p{#1}}
\newcolumntype{J}[1]{>{\arraybackslash}p{#1}}

\renewcommand{\theFancyVerbLine}{\textsuperscript{\arabic{FancyVerbLine}}}
\renewcommand{\theFancyVerbLine}{\textsubscript{\arabic{FancyVerbLine}}}
\renewcommand{\theFancyVerbLine}{\varepsilon}

\def\cD{{\cal D}}

\begin{frontmatter}

\title{Evaluation of Feynman integrals with arbitrary complex masses via series expansions}

\author[a,b]{Tommaso Armadillo}\ead{tommaso.armadillo@uclouvain.be}
\author[c,d]{Roberto Bonciani}\ead{roberto.bonciani@roma1.infn.it}
\author[b,e]{Simone Devoto}\ead{simone.devoto@unimi.it}
\author[b,e,f]{Narayan Rana}\ead{narayan@iitk.ac.in}
\author[b,e]{Alessandro Vicini}\ead{alessandro.vicini@mi.infn.it}

\address[a]{Centre for Cosmology, Particle Physics and Phenomenology (CP3),
Universit\'{e} Catholique de Louvain, Chemin du Cyclotron, B-1348 Louvain la Neuve, Belgium}
\address[b]{Dipartimento di Fisica ``Aldo Pontremoli'', University of Milano, Via Celoria 16, I-20133 Milano, Italy}
\address[c]{Dipartimento di Fisica, Universit\`{a} di Roma ``La Sapienza'', Piazzale Aldo Moro 5, I-00185 Roma, Italy}
\address[d]{INFN Sezione di Roma, Piazzale Aldo Moro 5, I-00185 Roma, Italy}
\address[e]{INFN Sezione di Milano, Via Celoria 16, I-20133 Milano, Italy}
\address[f]{Department of Physics, Indian Institute of Technology Kanpur, 208016 Kanpur, India}


\begin{abstract}
  We present an algorithm to evaluate multiloop Feynman integrals
  with an arbitrary number of internal massive lines,
  with the masses being in general complex-valued, and its implementation in the \textsc{Mathematica} package \textsc{SeaSyde}.
  The implementation solves by series expansions
  the system of differential equations
  satisfied by the Master Integrals.
  At variance with respect to other existing codes,
  the analytical continuation of the solution is performed
  in the complex plane associated to each kinematical invariant.
  We present the results of the evaluation of the Master Integrals
  relevant for the NNLO QCD-EW corrections to
  the neutral-current Drell-Yan processes.
\end{abstract}

\end{frontmatter}

\begin{small}
\noindent
{\em Program Title:}  {\textsc SeaSyde}                                        \\
{\em CPC Library link to program files:} (to be added by Technical Editor) \\
{\em Developer's repository link:}   \url{https://github.com/TommasoArmadillo/SeaSyde}     \\
{\em Code Ocean capsule:} (to be added by Technical Editor)\\
{\em Licensing provisions(please choose one):} GPLv3  \\
{\em Programming language:}   {\textsc  Mathematica  }                                \\
{\em Supplementary material:}                                 \\
{\em Nature of problem:}  Solution of multi-loop Feynman integrals with an arbitrary number of internal complex-valued massive lines. \\
{\em Solution method:}  Solution via series expansions of the systems of differential equations satisfied by the loop integrals.
Analytic continuation of the solution to any point of the kinematic variables. The solution is transported from the initial to the final point, by moving in the complex plane associated to each kinematic variable. In multi-variable problems, the transport is computed one variable at the time.    \\
   \\

\end{small}

\section{Introduction}
\label{sec:intro}

The high quality and precision of the large amount of data collected by the experiments at the CERN LHC calls for a corresponding theoretical effort, to interpret the results in a meaningful way.
The evaluation of the radiative corrections at second and higher orders in the perturbative coupling constant expansion has become a need.
These challenging calculations require the solution of multiloop Feynman integrals, which are in several cases still unknown.
The complexity of these problems grows with the number of energy scales: independent kinematical invariants and masses. The latter might be complex valued, as it is the case for unstable particles like the gauge and the Higgs bosons or the top quark.
The gauge symmetry of the scattering amplitude leads to non-trivial cancellations among the different terms associated to the Feynman diagrams and possible problems of precision loss must be kept under control. The analytical knowledge of the solution of the Feynman integrals is thus desirable, to have an arbitrary number of digits available. When this is not the case, then different approaches might be useful.
Comprehensive reviews on the Feynman integrals, covering the various strategies can be found in Refs.\cite{Heinrich:2020ybq,Weinzierl:2022eaz,Bourjaily:2022bwx,Abreu:2022mfk,Blumlein:2022zkr}.

It is difficult to design  a general algorithm which allows the evaluation of an arbitrary Feynman integral, because of the presence of singular points and branch cuts, connected to the physical thresholds and pseudo-thresholds of the diagram. For this reason a purely numerical approach \cite{Smirnov:2015mct,Borowka:2017idc} can encounter problems whenever an integrable singularity is present. Isolating the singular points and working out a dedicated analytical rearrangement of the integrand function may yield the successful evaluation of the integral, decomposed in the sum of several regular terms.
Whenever the identification of the singular structure or the following rearrangement can not be performed, then numerical instabilities are expected. The analytical solution is clearly preferable in all such cases, because the singular behaviour can be exposed and arbitrary precision can be achieved everywhere else.

We have analytical control over the solution of an integral, if it can be expanded as a power series at every point of its domain.
We dub analytical solutions those which can be expressed in closed form as a combination of elementary and special functions, whose power expansion is known. We call semi-analytical solutions those which can only be represented as power expansions, without additional functional relations (the latter are typically available in the closed form case).

Beside the different strategies to solve the problem of integration of the Feynman integrals, the differential equations approach \cite{Kotikov:1990kg,Kotikov:1991pm,Bern:1993kr,Remiddi:1997ny,Gehrmann:1999as,Argeri:2007up,Henn:2013pwa,Henn:2014qga} has become very popular, covering a large number of cases relevant in multi-loop calculations.
The Feynman integrals can be written as a combination of basic integrals called Master Integrals (MIs) and the latter satisfy, in fact, systems of first order linear differential equations. These systems can be solved, in several relevant cases, in closed form: the system is expanded in the dimensional regularisation parameter $\varepsilon=(4-d)/2$, where $d$ is the dimension of the space-time and, order by order in $\varepsilon$, the solutions are expressed in terms of a suitable functional basis. A first class of problems is represented by the systems that admit solutions expressed in terms of 
generalised polylogarithmic functions \cite{Goncharov:polylog,Goncharov:1998kja,Goncharov:2001iea,Remiddi:1999ew}.
%
%
The algebraic properties of these functions allow to simplify the final expression of the scattering amplitude. Moreover, their numerical evaluation is under control and provided by general public routines \cite{Gehrmann:2001pz,Gehrmann:2001jv,Bauer:2000cp,Vollinga:2004sn,Bonciani:2010ms,Naterop:2019xaf}.
%
%
Recently, another class of problems became relevant for phenomenological applications, namely cases in which the systems admit solutions expressed in terms of more general special functions, e.g. of elliptic kind \cite{Adams:2016xah,Remiddi:2017har,Broedel:2017kkb,Broedel:2017siw,Broedel:2018iwv,Ablinger:2017bjx,Bourjaily:2022bwx}. Also in this case, the properties of the functions under consideration allow for a simplification in the final expressions and a control on the complexity of the analytic formulae. The numerical evaluation of such functions is less general, but nevertheless possible using power expansion representations \cite{Walden:2020odh}. 
Cases in which a closed form instead is not known or not suitable for power expansions need to be treated using a different strategy.

We can thus summarise that the study of a multi-loop 
Feynman integral leads to two distinct cases: $i)$ the solution is achievable in terms of special functions; $ii)$ the closed form solution is not available.
We focus in this paper on the second case and consider the recent developments of semi-analytical algorithms to solve via series expansion the systems of differential equations satisfied by a Feynman integral
\footnote{
Different numerical approaches for the solution of differential equations have been proposed in Refs.~\cite{Czakon:2008zk,Mandal:2018cdj}.
}.
The solution is first written as a power series with unknown coefficients and is inserted in the differential equations; an infinite number of algebraic equations for the unknown coefficients can be written and recursively solved. The solution is thus known inside the convergence radius of the series expansion. The extension to different regions can be achieved via analytic continuation.
 This method was applied to the evaluation of one-dimensional problems (sunrise and vertex corrections, in which the MIs depend upon a single dimensionless variable) \cite{Pozzorini:2005ff,Aglietti:2007as,Lee:2017qql,Lee:2018ojn,Bonciani:2018uvv,Fael:2021kyg,Fael:2022rgm}. Then it was generalised to multi-dimensional problems in \cite{Moriello:2019yhu}. Recently a {\sc Mathematica} package, called {\sc DiffExp}, has been presented \cite{Hidding:2020ytt}, allowing the solution of an arbitrary system of differential equations, provided that the boundary conditions and the necessary prescriptions for the analytic continuation are known. This package has been successfully applied to several different problems \cite{Bonciani:2019jyb,Frellesvig:2019byn,Abreu:2020jxa,Dubovyk:2022frj,Bonciani:2021zzf,Becchetti:2020wof,Becchetti:2021axs}, always considering the case of real-valued masses for the particles running inside the loops.

 The electroweak (EW) precision physics program at the LHC and future colliders requires the evaluation of EW radiative corrections which involve the exchange of $W$, $Z$, and the Higgs bosons. The latter are unstable particles and the gauge invariant definition of their masses can be achieved in the Complex-Mass-Scheme (CMS)\cite{Denner:1999gp,Denner:2005fg}, identifying the complex-valued pole of their propagator. Precise predictions for the production of the top quark require the inclusion of off-shellness effects including the top decay width.
 In this paper we present a package which is able to deal with complex-valued masses in the internal lines of the Feynman integral. We develop an original independent algorithm to implement the analytic continuation.

The paper is organised as follows.
In Section~\ref{sec:series} we provide a pedagogical introduction to the solution of a system of first order linear differential equations by series expansions, with a particular focus on how the analytic continuation of the result can be safely obtained to an arbitrary point in the complex plane. 
We outline the implementation in the \textsc{Mathematica} package \textsc{SeaSyde}  (Series Expansion Approach for SYstems of Differen-
tial Equations) of the algorithm that solves a system of differential equations, for arbitrary complex-valued kinematical variables and we illustrate the different computational strategies available.
In Section~\ref{sec:results}, as a first practical application of the algorithm, we discuss the solution of the MIs needed to evaluate the two-loop QCD-EW virtual corrections to the neutral-current Drell-Yan processes \cite{Bonciani:2021zzf,Armadillo:2022bgm}.
The detailed expressions and evaluation properties of the integrals used in Ref.\cite{Armadillo:2022bgm} constitute an original result of this paper.
Finally, in Section~\ref{sec:conclusion} we draw our conclusions.

The latest version of the \textsc{Mathematica} package \textsc{SeaSyde} can be downloaded from \url{https://github.com/TommasoArmadillo/SeaSyde}, while its full documentation for Version 1.0 is provided in~\ref{app:packagedoc}.

\section{The solution algorithm}
\label{sec:series}
The solution of a system of linear differential equations by series expansions is well known in the mathematical literature.
In this Section we present some basic definitions and a pedagogical introduction to the procedure, and we show how it can be turned into an algorithm and applied to the specific case of Feynman integrals.

\subsection{Solving differential equations by series expansion}
\label{sec:example}
The general method to obtain the solution as a series expansion can be illustrated with a simple example.
Let us consider the differential equation
\be
\left\{
\begin{array}{l}
f'(x) +\frac{1}{x^2-4x+5}f(x)=\frac{1}{x+2}\\
f(0)=1
\end{array}\;;
\right.
\label{eq:exampleeqdiff}
\ee
with $x$ being a real variable.

In order to solve the homogeneous equation we use the well-known \textit{Frobenius method}. To this aim we introduce $f_{hom}(x)=x^r \sum_{k=0}^\infty c_k x^k$ as the expansion of the homogeneous solution around $x=0$, with $c_k$ being arbitrary coefficients that we need to determine.
By replacing $f_{hom}(x)$ with its power series in the associated homogeneous differential equation and by collecting all the terms with the same power of $x$, we obtain an infinite set of algebraic equations in the unknowns $c_k$.
In our example, Eq.~(\ref{eq:exampleeqdiff}) leads to:

\begin{equation}
\label{eq:example_syst}
\left\{
\begin{array}{l}
{r}{c_0}=0
\\
\frac{1}5 c_0+c_1(r+1)=0
\\
\frac 4{25} c_0+\frac 15 c_1+c_2(2+r)=0
\\
\frac{11}{125}c_0+\frac{4}{25} c_1+\frac 15 c_2+c_3(3+r)=0
\\
\dots
\end{array}
\right.
\end{equation}

The equation associated to the lowest power of $x$ is called indicial equation, and it assigns the value of the exponent $r$.
In our case, we get ${r}{c_0}=0$, which corresponds to $r=0$.
The other equations determine all the $c_k$ but one; the latter can be chosen arbitrarily, and we set e.g. $c_0=5$.
We thus obtain the solution of the homogeneous differential equation associated to the one in Eq.~(\ref{eq:exampleeqdiff}), expressed as the following series expansion:
\be
f_{hom}(x)=
5- x -\frac3{10} x^2 - \frac{11}{150} x^3+...
\ee
A particular solution for the original problem can now be obtained by applying the variation of the constant method, where the inverse of the homogeneous solution, multiplied by the inhomogeneous term, is expanded about $x=0$ and easily integrated:
\be
f_{part}(x)=
f_{hom}(x)
\int_0^x dx'\, \frac{1}{(x'+2)}\, f^{-1}_{hom}(x')
=
\frac 12 x-\frac 7{40} x^2+\frac 2{75} x^3+...
\label{eq:ex1}
\ee
The general solution is finally given by 
\be
f(x)=f_{part}(x)+C f_{hom}(x)\;.
\ee
In order to satisfy the boundary condition $f(0)=1$ we set $C=1/5$.


\subsection{Singularities and branch cuts}
\label{sec:cuts}

The validity of a solution obtained as a series expansion, with the procedure outlined in Section~\ref{sec:example}, is limited by its convergence radius.
We now discuss how the latter is defined and how the solution can be extended, via analytic continuation, from its initial region of convergence to an external arbitrary point.

\begin{figure}[t]
\centering
\includegraphics[width=0.5 \textwidth]{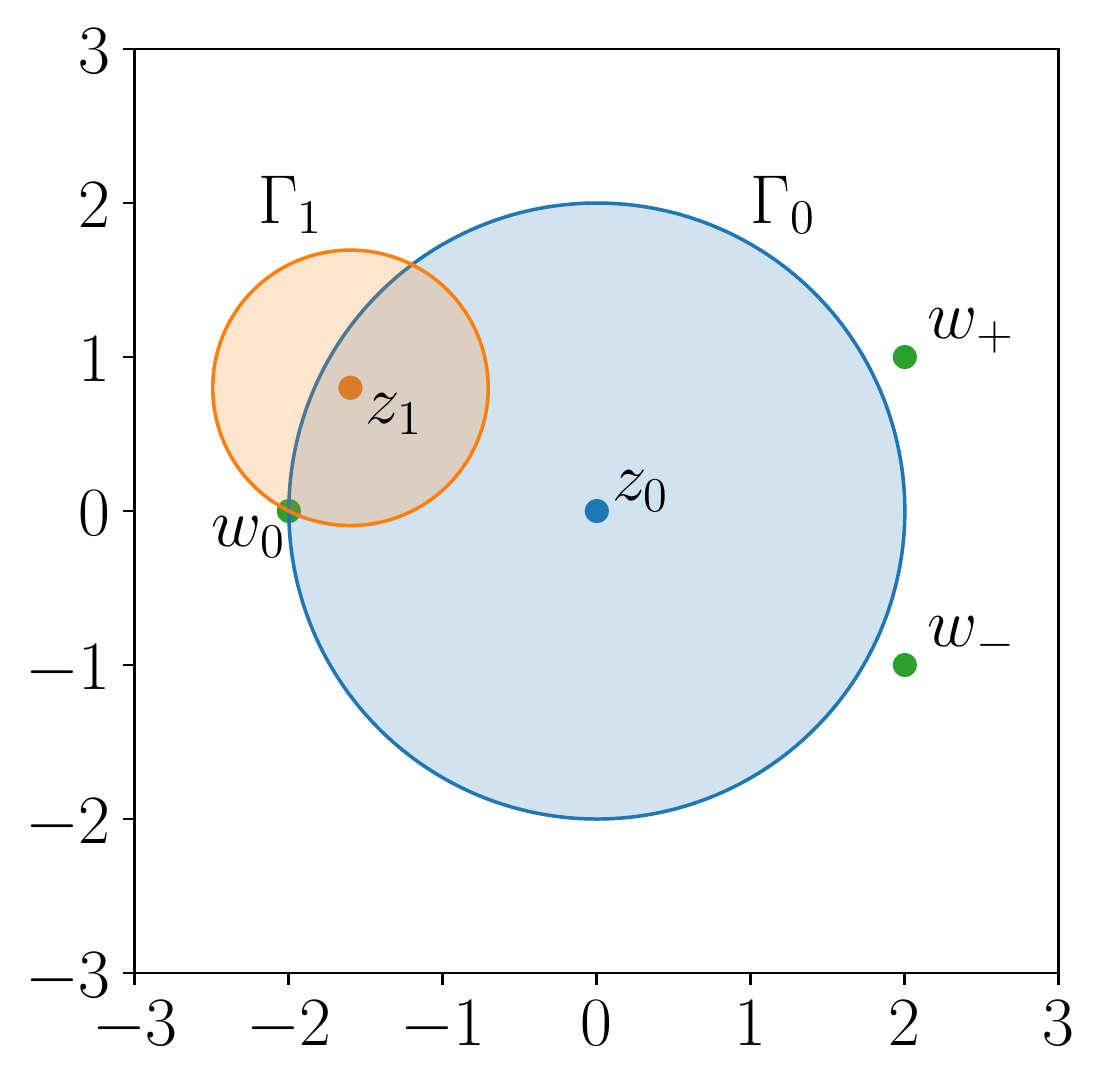}
\caption{\label{fig:analcont}
  Example of analytic continuation.}
\end{figure}

A solution written in series representation converges in the complex-$z$ plane inside a disc centered around the expansion point with a convergence radius limited by the closest singularity.
As an example, we consider again the first-order linear differential equation presented in Eq.~(\ref{eq:exampleeqdiff}), now as a function of a complex variable $z$.
By analysing the differential equation, we can identify three singularities: the factor $(z+2)$ in the denominator of the inhomogeneous part generates a pole in $z=w_0=-2$, while the factor $(z^2-4z+5)$ in the denominator of the homogeneous part implies the presence of two poles in $z=w_\pm=2\pm i$.
The solution presented in Eq.~(\ref{eq:ex1}) thus converges in the complex plane within a disc $\Gamma_0$ centered in $z_0=0$ with radius $2$, as illustrated by the blue circle in Figure~\ref{fig:analcont}.

We can analytically continue the solution into a new disc, centered at an arbitrary point $z_1$ internal to $\Gamma_0$. Also in this case, the convergence radius is defined by the closest singular point, as illustrated in Figure~\ref{fig:analcont} by the orange circle $\Gamma_1$.
It is possible to demonstrate that this procedure is unique.

In the case of simple poles in the homogeneous coefficient, we obtain a logarithmic behaviour of the solution, that requires some additional care in the analytic continuation.
The presence of logarithmic functions makes the solution multi-valued: it can be made single-valued by adding cuts in the complex plane, thus specifying the Riemann sheet where the function is evaluated.
This feature is clearly not present in the power series representation, but must be introduced to allow for a physical interpretation of the results.
To take it into account, we need to define in an unambiguous way the cut associated to each singularity that shows a logarithmic behaviour: their presence will then affect the process of analytic expansion.

In our example, we now consider a logarithmic behaviour in the points $w_0$, $w_\pm$, and we choose as the branch-cuts the horizontal lines parallel to the real axis that go from the singular point to $-\infty$.
In the following discussion we will always assume this convention,
which represents the standard choice for logarithmic branch cuts\footnote{This conventions is used, for instance, in {\sc Mathematica} by the function $\tt{Log[z]}$, for complex values of the variable $\tt{z}$.}.

\begin{figure}[t]
\includegraphics[width=0.49\textwidth]{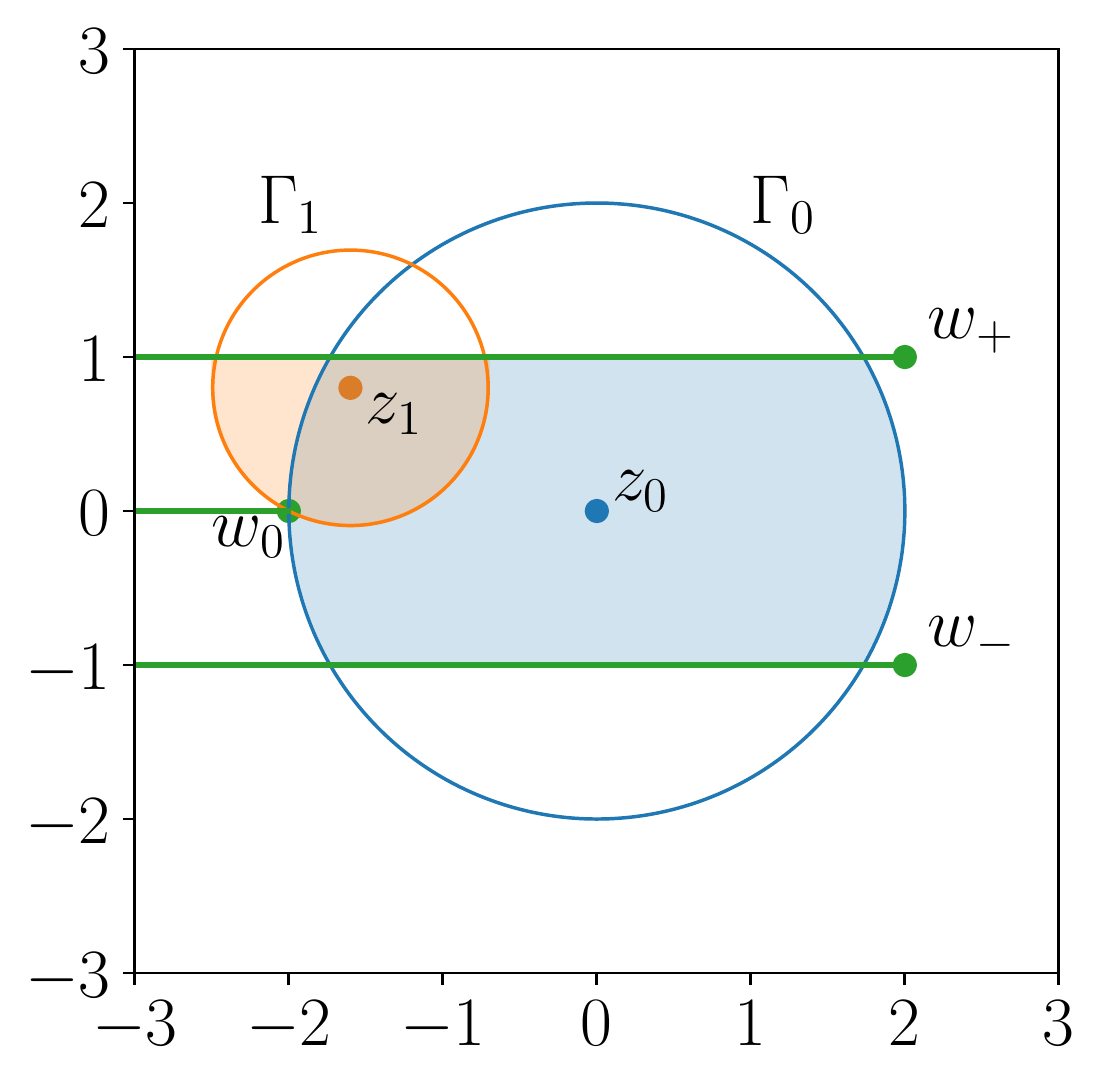}
\includegraphics[width=0.49\textwidth]{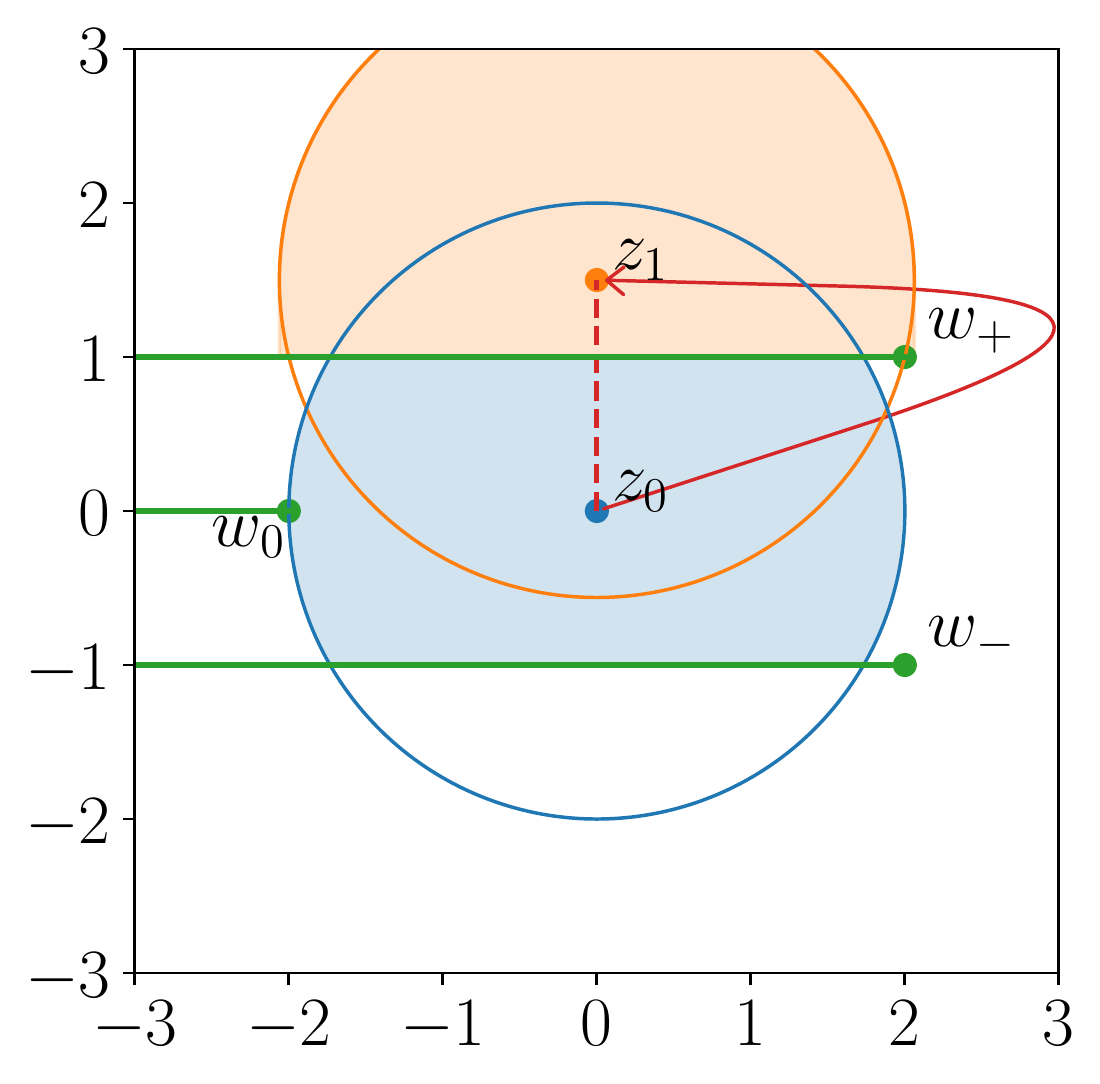}
\caption{\label{fig:cuts}
 Example of the effect of branch-cuts on the convergence of the expanded solution: reduced convergence area (left) and different path for the analytic continuation (right).}
\end{figure}

The first consequence of the presence of branch-cuts in the procedure of analytic continuation is on the area of convergence of the series expansion.
While the proximity with a branch-cut does not modify the convergence radius of the series itself, it limits the area of the disc where the expansion converges to the desired value: 
once the cut is crossed, the series is evaluated on a Riemann sheet different than the one that we have chosen with the cuts, and thus the result can not be considered as the solution of the problem in the assigned domain.
This effect is illustrated on the left panel of Figure~\ref{fig:cuts}, that shows how the convergence discs $\Gamma_0$, $\Gamma_1$ introduced in Figure~\ref{fig:analcont} are modified once the branch cuts are inserted: while the discs themselves do not shrink, only the reduced highlighted area provides the correct evaluation of the solution.

The second consequence is on the path that is necessary to follow in order to extend the solution to an arbitrary point in the complex plane. The path has to be chosen in such a way that it does not cross any branch cut. This is illustrated on the right panel of Figure~\ref{fig:cuts}, where the dotted path from $z=0$ to $z_1$ is now forbidden by the presence of the branch cut corresponding to the pole $w_+$. The cut needs to be avoided with a path that goes around to the right of the singularity, as shown by the solid line. For the sake of readability, we have not drawn all the discs along the path, representing the intermediate steps of the analytic continuation.

\subsection{Choice of the path}
\label{subsec:path}
In the {\sc Mathematica} package {\sc SeaSyde}, we use the same convention for the branch-cuts as the one we introduced in the previous section, with branch-cuts parallel to the real axis and going from the singularities to $-\infty$.
As it has been shown in Section~\ref{sec:cuts}, with this choice the value of the solution in each point of the complex plane is unambiguously defined, as long as the path we use for the analytic continuation does not cross any branch-cut.
In the following, we briefly describe how such path can be defined when connecting two arbitrary points on the complex plane and the algorithm for its determination implemented within the package.
The analytic continuation described in the previous Section, moving in the complex plane of each kinematical invariant, allows to treat exactly in the same way the cases of real- and complex-valued masses. This represents the main remark of this paper.
\begin{figure}[t]
\includegraphics[width=0.49\textwidth]{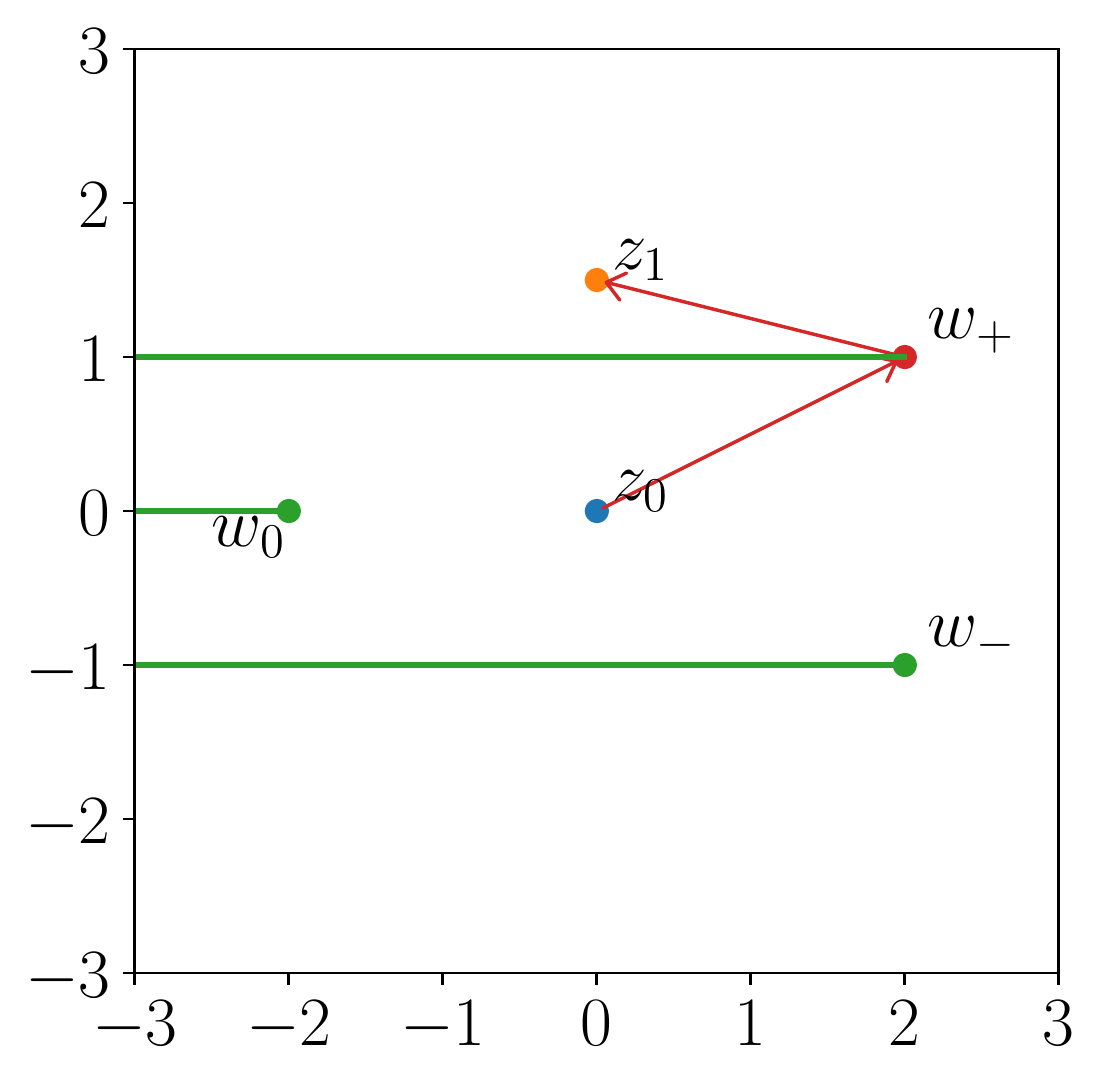}
\includegraphics[width=0.49\textwidth]{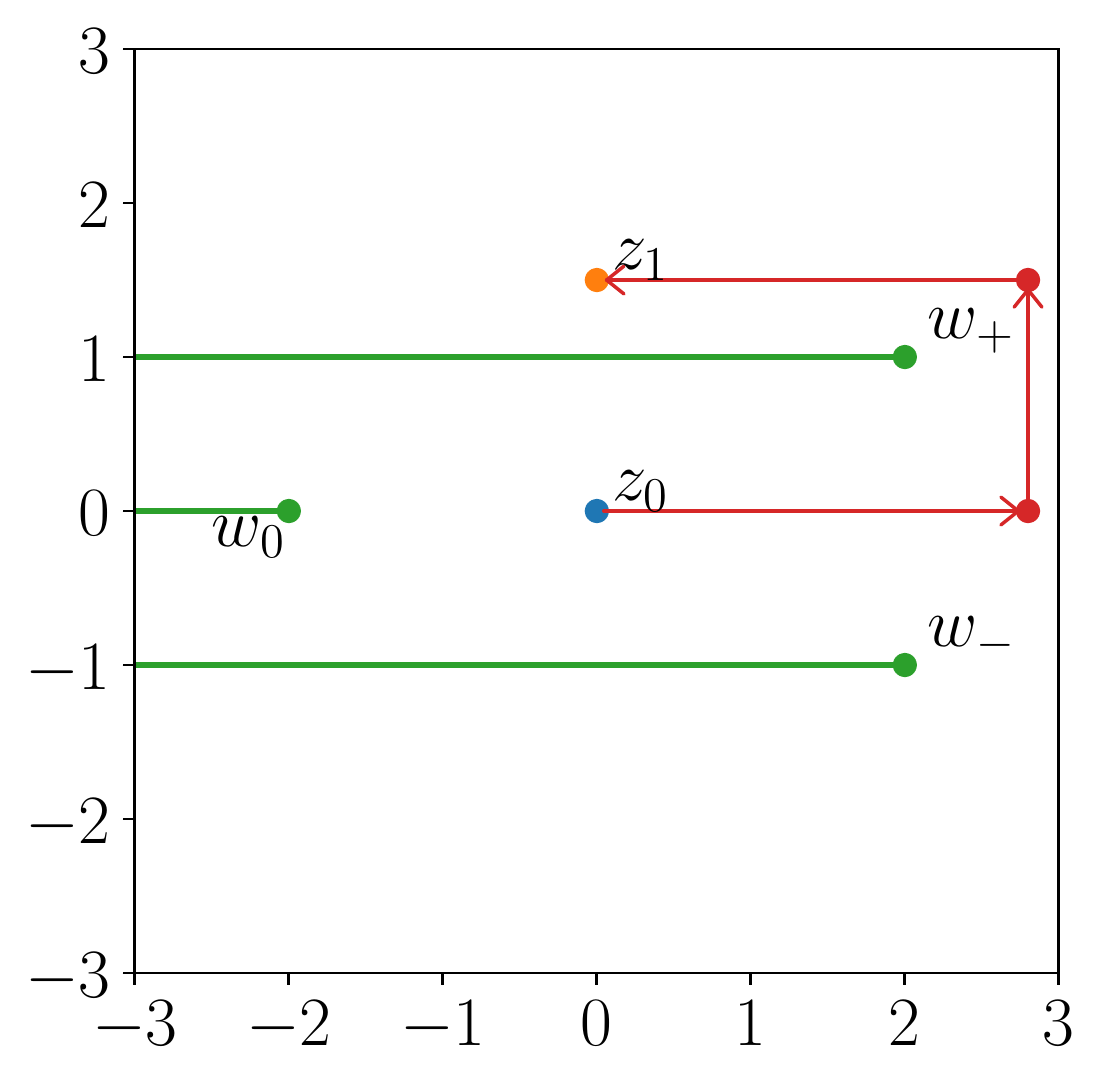}
\caption{\label{fig:path_logVStaylor}
 The two possible approaches in the definition of the path: the one requiring a logarithmic expansion (left) and the one only relying on the Taylor expansion (right).}
\end{figure}

There are two main approaches that can be used while defining the path: we can avoid the singularities or we can use them as an expansion point.
This two possibilities are illustrated in Figure~\ref{fig:path_logVStaylor}, where we consider once again the pole structure given by Eq.~(\ref{eq:exampleeqdiff}), and we aim to move the boundary conditions from the point $z_0$ to the point $z_1$.
By choosing the singularity in $w_+$ as one of the possible expansion points, as shown in the left panel of Figure~\ref{fig:path_logVStaylor}, we obtain a logarithmic expansion around $w_+$, while if we avoid the singularity, as shown in the right panel of Figure~\ref{fig:path_logVStaylor}, we only rely on the ordinary Taylor expansion.
While leading exactly to the same result, the two approaches have different advantages and disadvantages.
By using the logarithmic expansion, we can expect a larger radius of convergence of the series, since the latter is not affected anymore by the presence of the pole upon which we are expanding: this can effectively reduce the number of steps required to reach the point of interest.
Furthermore, the presence of the branch-cut, starting from the singularity we are expanding upon, is automatically taken into account by the explicit logarithms appearing in the series.
On the other hand, in the current implementation of {\sc SeaSyde}, the expansion on a singular point requires a longer evaluation time with respect to the ordinary Taylor expansion on a regular point of the complex plane.
An optimised choice between the two different approaches thus requires to carefully balance these two effects.
In the physical applications we have dealt with while testing the code, we have observed how the presence of multiple singularities close to each other does not allow for a significant improvement of the convergence radius while using the logarithmic expansion. This fact has led us to prefer, for our final implementation, the more regular behaviour of the ordinary Taylor expansion.

With this choice, the algorithm to determine the path is straightforwardly implemented.
Since all the branch-cuts are parallel to each other and in the same direction, it is always possible to connect two arbitrary points on the complex plane with a path that goes around to the right of the rightmost singularity that lays between them, as it is shown in the right panel of Figure~\ref{fig:path_logVStaylor}. 

\begin{figure}[t]
\includegraphics[width=0.49\textwidth]{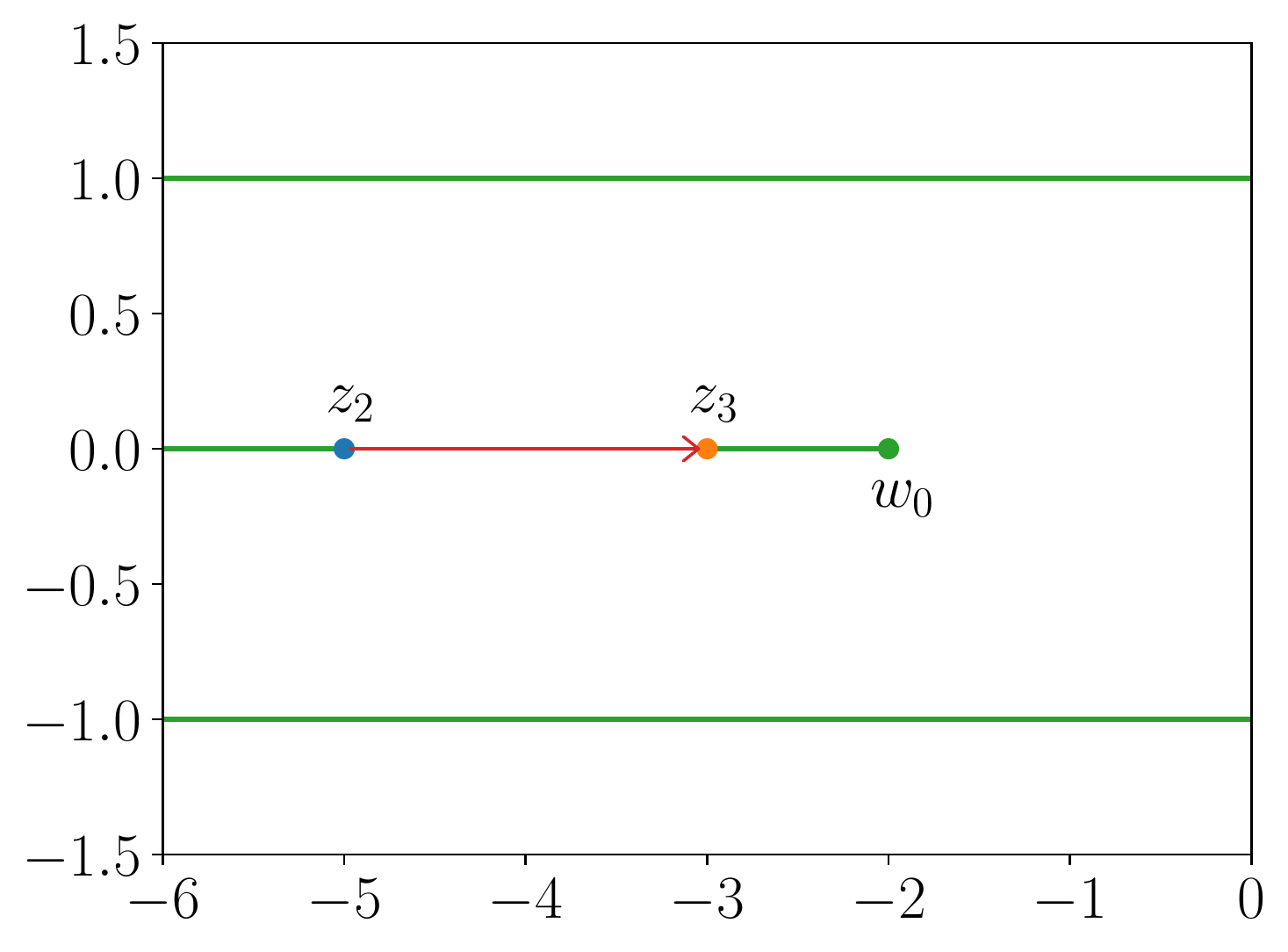}
\includegraphics[width=0.49\textwidth]{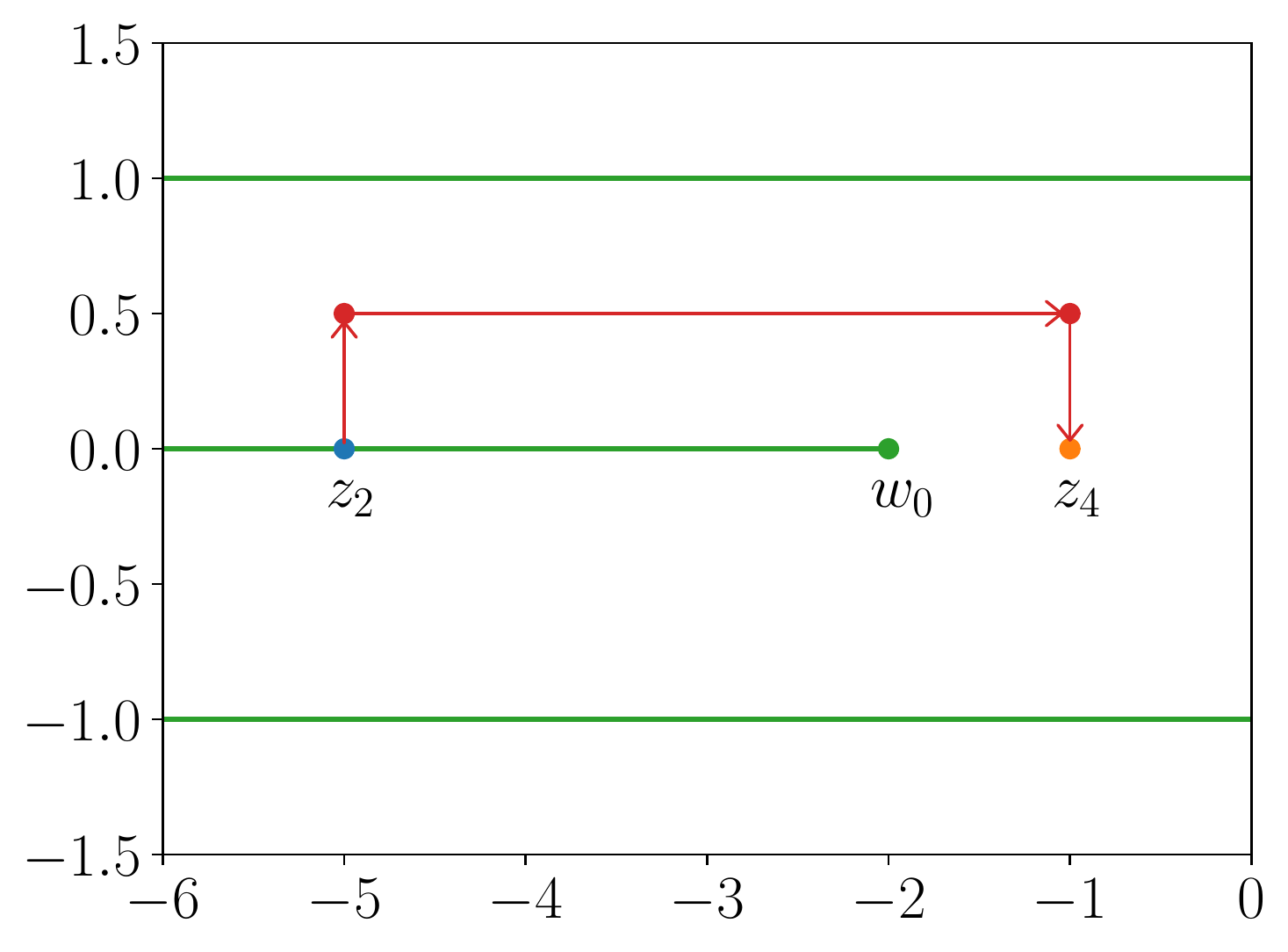}
\caption{\label{fig:realAxis}
Examples of two possible paths for linking two points laying on the same branch-cut: no singularities between them are present (left) or at least one is (right).}
\end{figure}

There is only one edge case that we need to discuss in more detail, that is the one in which the starting and ending point lay down on the same branch-cut, which is usually the real axis. In this case, indeed, there is an ambiguity, since it is not clear where the points are located with respect to the cut. This ambiguity can be discussed and solved in the light of the Feynman prescription for the particle propagators, which enforces the causality requirement of the theory.
If we consider, for the sake of simplicity, the propagator $\Pi(s)=i/(s-m^2+i\delta)$ of a massive scalar field, despite the kinematic invariant $s$ and the mass $m$ being real, the Feynman prescription shifts the first one to $s+i\delta$, i.e. in the upper half of the complex-$s$ plane.
Having a prescription which pushes the solution of the differential equation in one specific half of the complex-$z$ plane, removes any ambiguity associated with the evaluation of the function on the branching cut and allows us to uniquely determine the solution for any given Boundary Condition (BC).
From a practical point of view, we might encounter two different scenarios: without or with poles between the start and the end points. In the first one the choice of the path is trivial and we can move directly from one point to the other, because, by construction, we are never crossing the cut. We can see this in the left panel of Figure~\ref{fig:realAxis}. 
In the second one, instead, we could again either expand on the singularity or avoid it, as previously discussed. The choice we have made to avoid the singularities, however, forces us to adopt an horseshoe path, as depicted in the right panel of Figure~\ref{fig:realAxis}.
A prescription of $-i\delta$, on the contrary, would have implied an horseshoe in the lower half of the complex plane.
The prescription is not hard-coded within {\sc SeaSyde} and must be passed as an input parameter. As a consequence, the package can automatically perform the analytic continuation of the solution even if the user adopts different conventions.

\subsection{Solving a system of differential equations}
\label{sec:system}
The evaluation of a scattering amplitude requires the numerical calculation of a large set of Feynman integrals. Their number can range from a few, for one-loop corrections, to hundreds or thousands, for multi-loop ones. Fortunately not all of them are independent from each other.  Integration-by-Parts Identities \cite{Tkachov:1981wb,Chetyrkin:1981qh,Laporta:2001dd} allow us to express these integrals as a linear combination of MIs\footnote{Several computer implementations of the procedure are available \cite{Anastasiou:2004vj,Studerus:2009ye,vonManteuffel:2012np,Lee:2012cn,Lee:2013mka,Smirnov:2008iw,Smirnov:2014hma,Maierhoefer:2017hyi,Klappert:2020nbg}.}.

The computation of the MIs can be faced in different ways. One of these possibilities is by writing the system of first order linear differential equations with respect to the kinematical invariants $\mathbf{s}$, satisfied by the MIs. If $\vec{I}(\varepsilon,\mathbf{s})$ is a vector with the MIs of the problem under consideration as components, the systems can be expressed as follows:
\be
\label{eq:system}
\frac{\partial}{\partial s_\alpha} \vec{I}(\varepsilon,\mathbf{s})
= \mathbf{A}_\alpha(\varepsilon,\mathbf{s})\, \vec{I}(\varepsilon,\mathbf{s})\, ,
\ee
where $s_\alpha$ is one of the invariants, 
and $\mathbf{A}_\alpha(\varepsilon,\mathbf{s})$ is the matrix of the coefficients of the system, containing rational functions of all the invariants and depending upon the parameter $\varepsilon$. We look for a solution in form of a Laurent series in $\varepsilon$, $\vec{I}=\sum_{k=-N}^{\infty} \varepsilon^k \, \vec{I}^{(k)}$, where $N$ is the order of the maximal pole of the MIs. We substitute this form of the solution in Eq.~(\ref{eq:system}) and, order-by-order in $\varepsilon$, we study the resulting system of equations, starting from the one associated to the lowest power of the expansion in $\varepsilon$.

A particularly suitable situation is the one in which the choice of a basis of MIs yields systems of equations that, order-by-order in $\varepsilon$, are in triangular form. This allows for a straightforward solution of the problem. The solution of the system can proceed in cascade: we solve a first-order linear differential equation at a time, starting from the one that involves a single MI. The solution of this equation is plugged in the following one, that becomes a first-order non-homogeneous differential equation. Also the solution of the second equation is plugged, together with the solution of the first one, into the third equation, that in turn becomes a first-order non-homogeneous differential equation as well. We proceed in this way up to the complete solution of the system\footnote{This is the class of problems that can admit solutions in closed form, in terms of polylogarithmic functions.
The system simplifies further, if it is possible, by an appropriate change of basis, to cast it in ``canonical form'' \cite{Henn:2014qga}.}.  
Each first-order differential equation can be solved using the procedure outlined in section~\ref{sec:example}, i.e. in a semi-analytic way.



However, the general situation is the one in which we face problems that are described by systems that include subsets of equations that are coupled\footnote{This means that we cannot solve just first-order linear differential equations, but we must solve, in principle, also higher-order differential equations, equivalent to the coupled subset under consideration.} at each order in $\varepsilon$. In these cases, the strategy that {\sc SeaSyde} uses to solve the systems is the following. The equations that are in triangular form are solved in cascade (as described above), while the subsets of coupled equations are solved expanding the corresponding masters in power series, with unknown coefficients. These coefficients are then fixed by the solution of linear systems of equations, constrained by the solution of the system of differential equations.

The {\sc Mathematica} package {\sc SeaSyde}, in its current implementation, can solve systems both in triangular and in coupled forms, covering the entire (current) set of possibilities encountered so far in the evaluation of Feynman diagrams for collider processes.


The systems of differential equations for a fixed order in $\varepsilon$ depend on multiple kinematic variables. We solve the problem by analysing only one of them at a time. 
In doing so, the dependence on all the other variables is parametric, and we reduce the problem to a system of differential equations in one variable.
The strategy adopted in  the package {\sc SeaSyde} is the following. Suppose that we are dealing with a problem depending upon two dimensionless variables, $x$ and $y$ . We ask {\sc SeaSyde} to evolve the solution from the point in which we fixed the boundary conditions, say $(x_0,y_0)$, to the point $(x_1,y_1)$. The solution of the problem will proceed solving the system in $x$, with a fixed value of $y=y_0$, from $x_0$ to $x_1$ and then solving the system in $y$ with $x=x_1$ from $y=y_0$ to $y=y_1$.
Integrating in $x$, at fixed $y=y_0$, the eventual cuts will depend on $y_0$. However, this value is fixed and then the cuts in the complex $x$ plane are given. To overcome the cuts, {\sc SeaSyde} uses the procedure outlined in section~\ref{sec:example}.
The same situation holds when we integrate in $y$ at fixed $x=x_1$. Now the eventual cuts will be in the complex $y$ plane and they will depend upon $x_1$, which is a fixed value. This means that also the cuts in $y$ are now given and {\sc SeaSyde} is able to overcome them with a path that moves at the right of the right-most branching point in $y$ (section~\ref{sec:example}).


We can start to solve the system from the lowest order in the $\varepsilon$ expansion and work our way up to the desired order.

\section{Two-loop MIs for the neutral-current Drell-Yan}
\label{sec:results}

As a first practical application of the algorithm discussed in section \ref{sec:series}, we provide the results for the evaluation of the MIs relevant for the mixed QCD-EW corrections to the neutral-current Drell-Yan process, recently presented in Ref \cite{Armadillo:2022bgm}. 
We focus on the integrals numbered 32-36 in Ref.~\cite{Bonciani:2016ypc},
because their complexity offers an interesting case to study the performances of our new package. The solution proposed in Ref.~\cite{Bonciani:2016ypc} in terms of Chen-Goncharov repeated integrals has severe limitations in the numerical evaluation in the physical region, and a semi-analytical approach represents an interesting alternative.

We first introduce the problem, then we show a comparison between the solution obtained with real and complex internal masses, and, finally, we discuss the numerical and time performances of our code. 

\subsection{Master integrals for the mixed EW-QCD corrections to the neutral-current Drell-Yan}

The NNLO mixed QCD-EW corrections to the neutral current Drell-Yan have been calculated recently in Ref. \cite{Armadillo:2022bgm}.
In the complete amplitude 204 MIs appear and they can be evaluated using the method of differential equations. 

These integrals can be classified into three different groups \cite{Bonciani:2016ypc}, according to the number of their internal massive lines: zero, one or two. 
The four-point functions corresponding to the first two sets are known in closed analytic form in terms of generalized polylogarithms (GPLs).
The last group
is 
the most challenging due to the presence of different kinematic scales. 
In particular 36 MIs belong to this set.
For 31 of them, a solution in terms of GPLs is known. 
The remaining 5 have been formally solved in \cite{Bonciani:2016ypc} using a formal expression in terms of Chen-Goncharov integrals \cite{Chen:1977oja}.
These MIs belong to the following integral family (see the classification in \cite{Armadillo:2022bgm}):
\be
\{ \cD_1, \cD_2-m_V^2, \cD_{12}, \cD_{1;1}, \cD_{2;1}, \cD_{1;12}, \cD_{2;12}-m_V^2, \cD_{1;3}, \cD_{2;3} \}
\ee
where $V$ can be either $W$ or $Z$ and $m^2_V$ is the mass-squared of the $V$ boson. We defined $\cD$s to be:
\be
    \cD_{i} = k_{i}^2,\;
    \cD_{ij} = (k_i-k_j)^2,\;
    \cD_{i;j} = (k_i-p_j)^2,\;
    \cD_{i;jl} = (k_i-p_j-p_l)^2,
\ee
where $p_i$ are the external momenta and $k_j$ the loop ones.
In particular the 5 MIs are:
\begin{align}
    &\text{Master 32}\;:\:\{0, 1, 1, 1, 0, 1, 1, 0, 1\} \quad &&\text{Master 33}\;:\:\{1, 1, 1, 1, 0, 1, 1, 0, 1\} \nonumber\\
    &\text{Master 34}\;:\:\{1, 1, 1, 1, -1, 1, 1, 0, 1\} \quad &&\text{Master 35}\;:\:\{1, 1, 1, 1, 0, 1, 1, -1, 1\}\nonumber\\
    &\text{Master 36}\;:\:\{1, 1, 1, 1, -1, 1, 1, -1, 1\}\;,
\label{eq:difficultMIs}
\end{align}
where we keep the numbering from Ref.~\cite{Armadillo:2022bgm}.\
It is possible to numerically evaluate the Chen iterated integrals in the Euclidean region, where the boundary conditions are imposed.
Their analytic continuation in the physical region, however, is non trivial and no standard technique to perform it is available. For this reason their computation in Ref.~\cite{Armadillo:2022bgm} has been performed by using a semi-analytical approach, that is by solving the relevant system of differential equations using a series expansion method as described in Section~\ref{sec:series}. 
By approaching the numerical evaluation of these MIs in such a semi-analytical way, we have to solve the complete system (36x36) of differential equations which includes the integrals from Eq.~(\ref{eq:difficultMIs}).

These MIs depend on two dimensionless kinematic invariants
$x$ and $y$:
\be
    s=(p_1+p_2)^2 \;,\qquad
    t=(p_1-p_3)^2 \;,\qquad
    x=-\frac{ s }{ M^2 }\;,\qquad
    y=-\frac{ t }{M^2 }\;,
    \label{eq:kinVariablesReal}
\ee
where $p_{1,2}$ are incoming momenta, $p_3$ is an outgoing momentum, $s,t$ are the Mandlestam invariant, and
$M$ is the mass of the vector boson.
Let us consider first the case of real-valued masses $M=m_V$, with $V=W,Z$.

Once we have chosen the point $(x,y)$
in which we would like to evaluate the solution, the latter can be obtained in a straightforward way
using \textsc{SeaSyde}:
\begin{Verbatim}[commandchars=\\\{\}] 
ConfigurationNCDY=\{                               
	EpsilonOrder->4,                               
	ExpansionOrder->50                                 
\};                                                      
UpdateConfiguration[ConfigurationNCDY];

SetSystemOfDifferentialEquation[SystemOfEquations, 
	        BoundaryConditions, MIs, \{x-I\textdelta,y+I\textdelta\}, pointBC]; 

TransportBoundaryConditions[ \{ tValue , sValue \} ]

SolutionValue[]
\end{Verbatim}

We use in this case a form of the MIs in which they are re-scaled by a power of $\varepsilon$, so that they do not contain explicit divergences in $\varepsilon$. 
Due to this re-scaling, in order to obtain the finite part for MI32-36 listed in Eq.~(\ref{eq:difficultMIs}) we need to keep 5 terms in the $\varepsilon$ expansion.
We have chosen to keep 50
terms in the expansion with respect to the kinematical variable, because it is a good compromise between execution time and precision; we will comment later 
how the number of terms affects the numerical precision. In \texttt{SetSystemOfDifferentialEquation} 
we are preparing the package for solving the system and, finally, using \texttt{Transport\-Boundary\-Conditions}
we are both solving the system and performing the analytic continuation, which let
us extend the solution from \texttt{pointBC}, where the boundary conditions are imposed, to \texttt{x=sValue} and 
\texttt{y=tValue}. At this point, by calling again the routine \texttt{Transport\-Boundary\-Conditions} we could transport the boundary conditions from \texttt{\{x=sValue, y=tValue\}} to 
\texttt{\{x=sValueNEW, y=tValueNEW\}} and so on. By repeating this procedure multiple times we can easily obtain a numerical grid for all the MIs. In \ref{app:packagedoc} we provide the full documentation of the package.

\subsection{Feynman integrals with complex-valued masses}

When dealing with intermediate unstable particles, such as $W$s and $Z$s, it is useful to perform the calculations in the CMS, in order to regularise the behaviour at the resonance while preserving the gauge invariance of the scattering amplitude.
To this end, we introduce the complex mass, defined as:
\be 
\label{eq:complex_mass}
\mu^2_V = m_V^2 - i \Gamma_V m_V,
\ee
where $\Gamma_V$, a real parameter, labels the decay rate of the boson.
The complex mass $\mu_V$ then replaces the real mass $m_V$ in all the steps of the computation.
From Eq.~(\ref{eq:kinVariablesReal}), in particular, we can observe that the
dimensionless kinematic variables $x$ and $y$ become complex-valued, once we set $M=\mu_V$.

The complex mass regularises the integrand function in the threshold region: indeed, all the propagators
\be
	\frac{1}{s-\mu_V^2+i\delta}
\ee
do not diverge for any real value of $s$, and, hence, in the kinematic region near the resonances, the MIs are smooth functions.

This fact let us really appreciate the discussion about the analytic continuation in the complex plane presented in Section~\ref{sec:series}:
thanks to the algorithm for performing the analytic continuation of the solution included in {\sc SeaSyde}, we are able to evaluate the 5 MIs of interest at every point of the physical region and for arbitrary complex variables, thus implementing the CMS in a straightforward way.

\begin{figure}[ht!]
    \centering
    \includegraphics[width=0.45\textwidth]{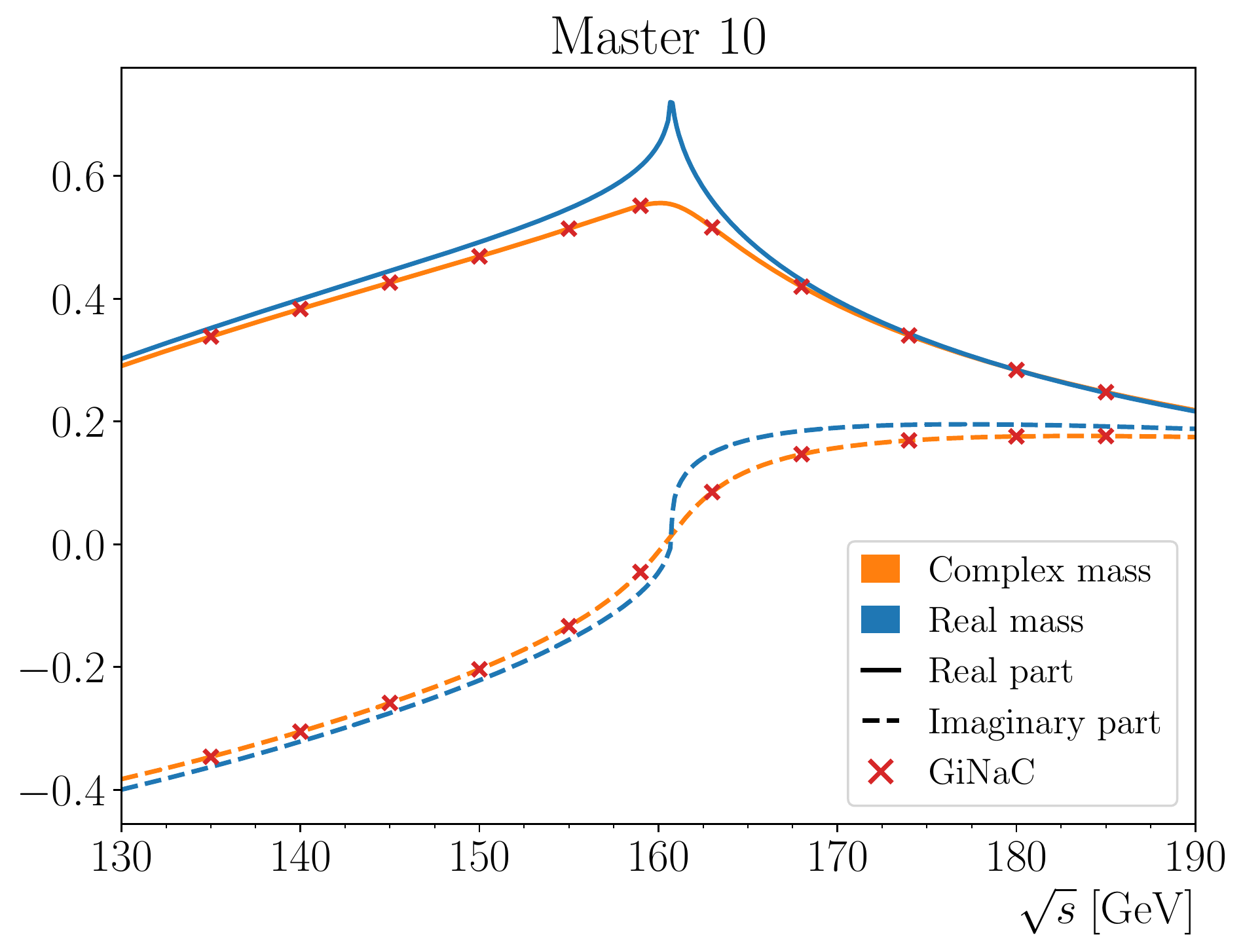}
    \includegraphics[width=0.45\textwidth]{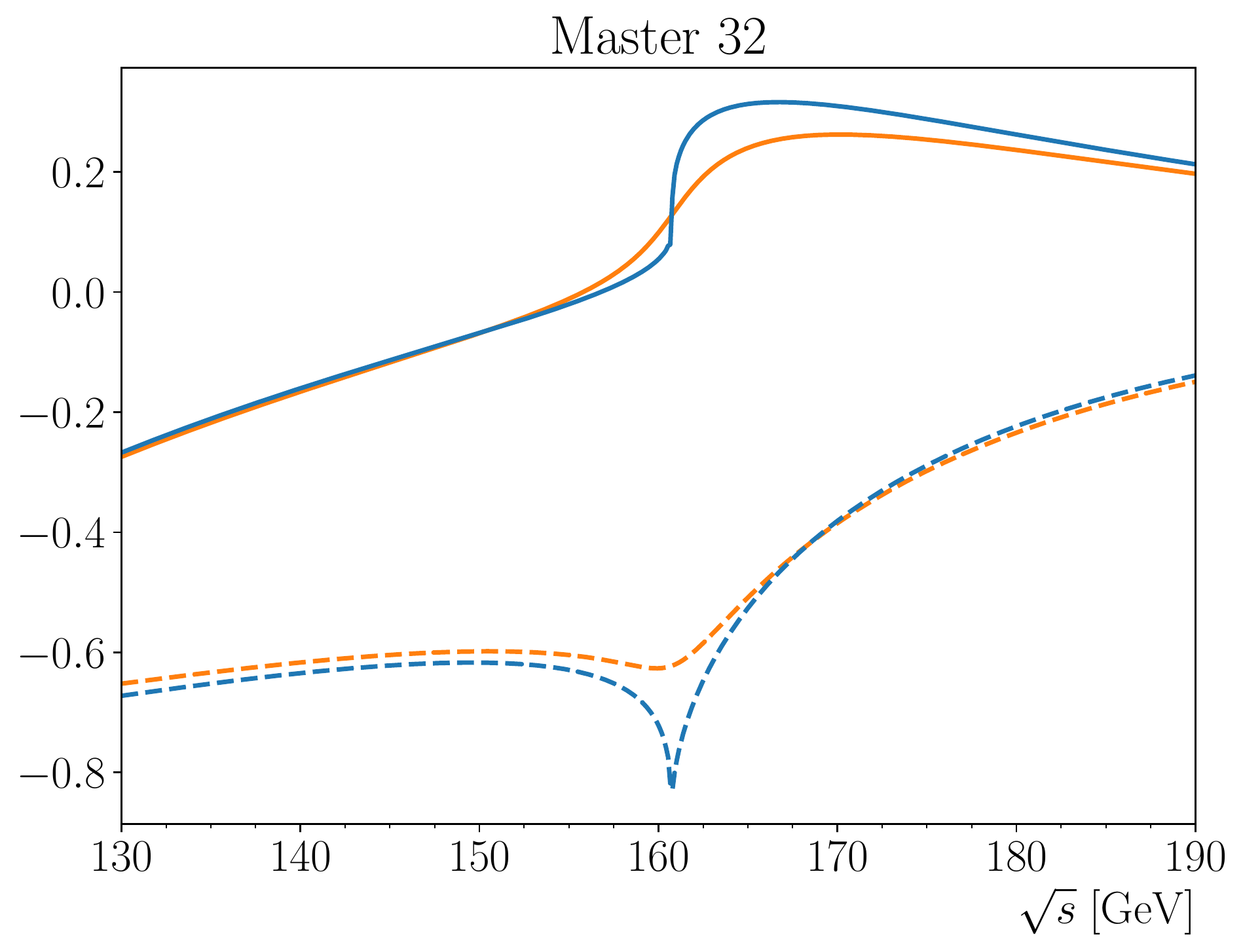}
    \includegraphics[width=0.45\textwidth]{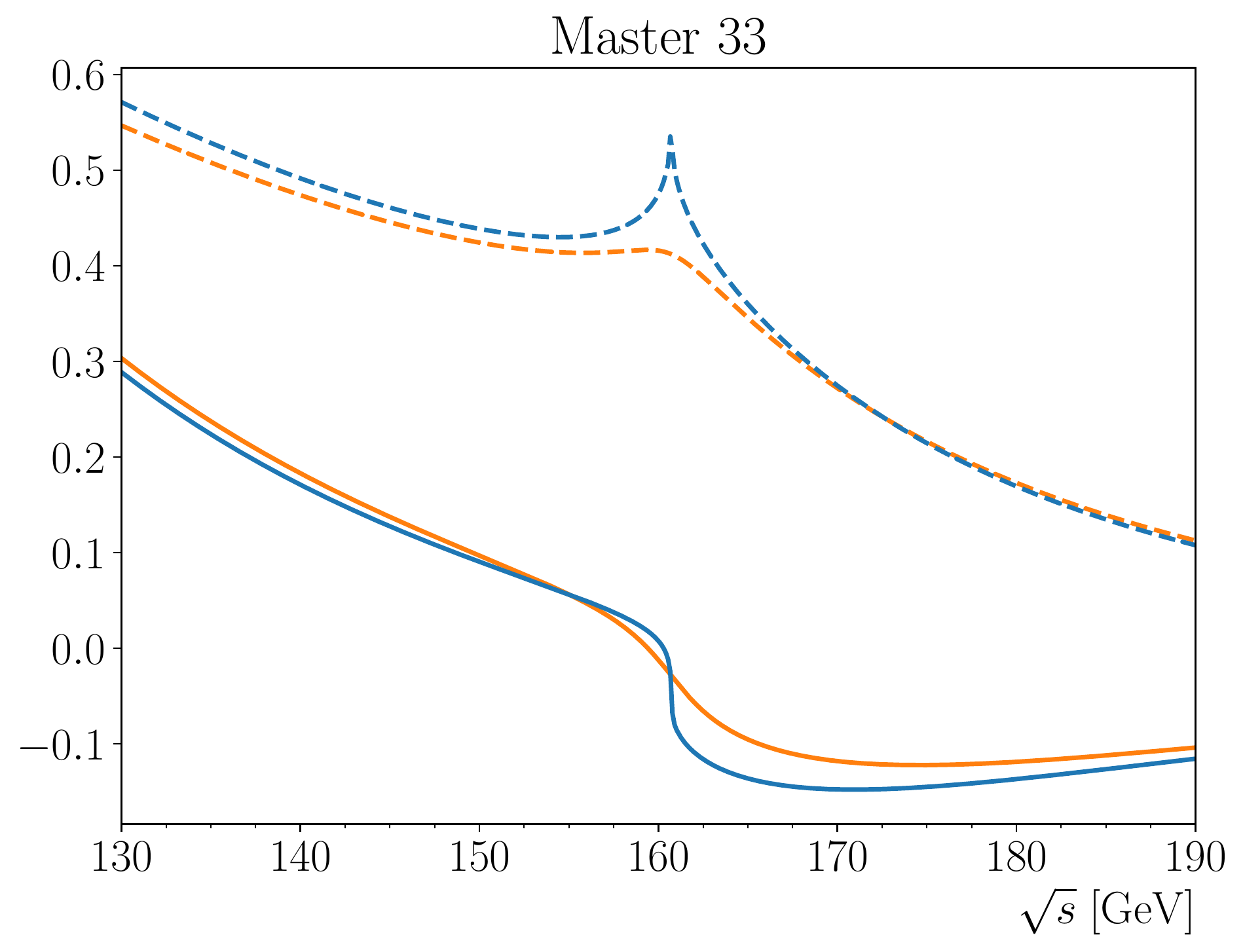}
    \includegraphics[width=0.45\textwidth]{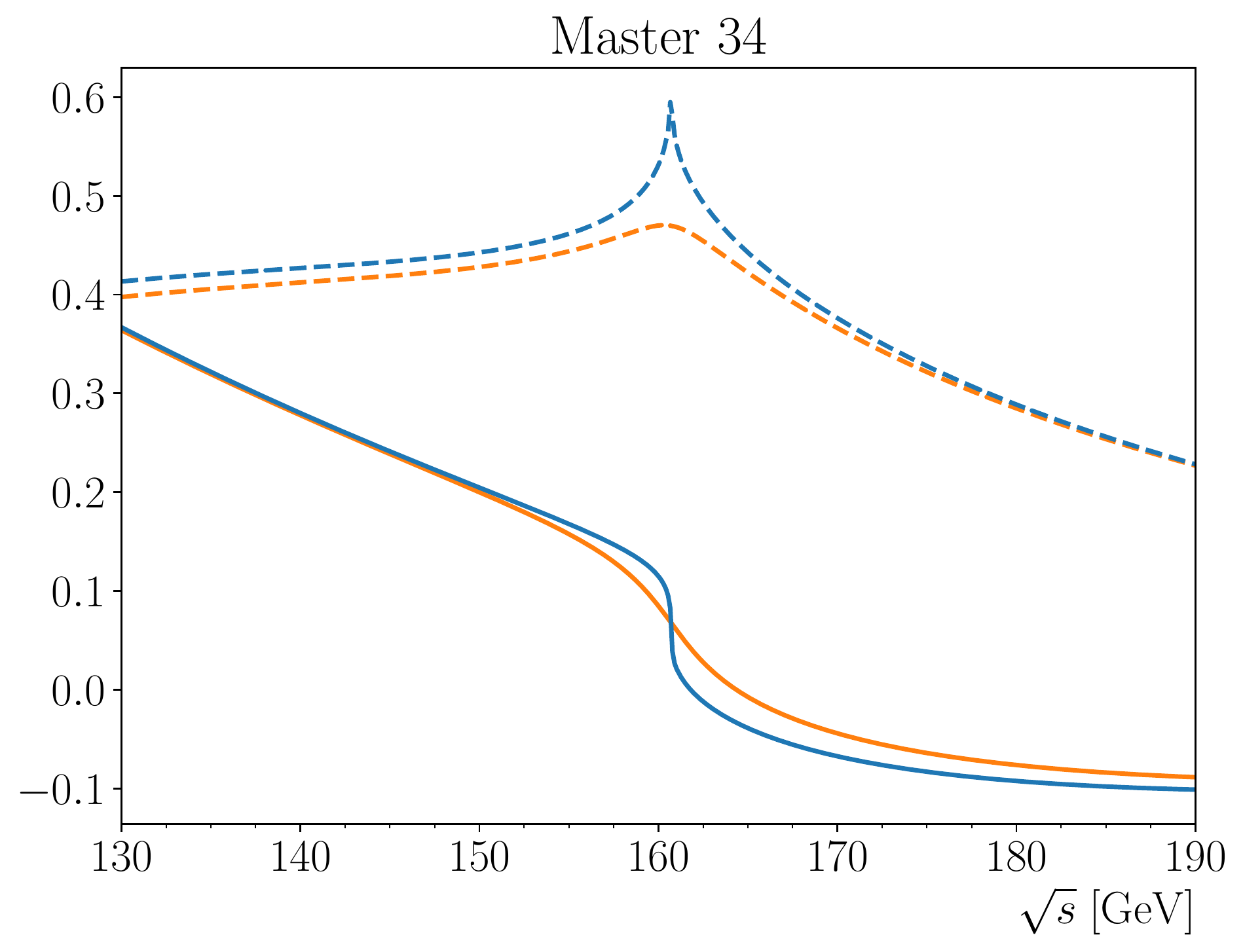}
    \includegraphics[width=0.45\textwidth]{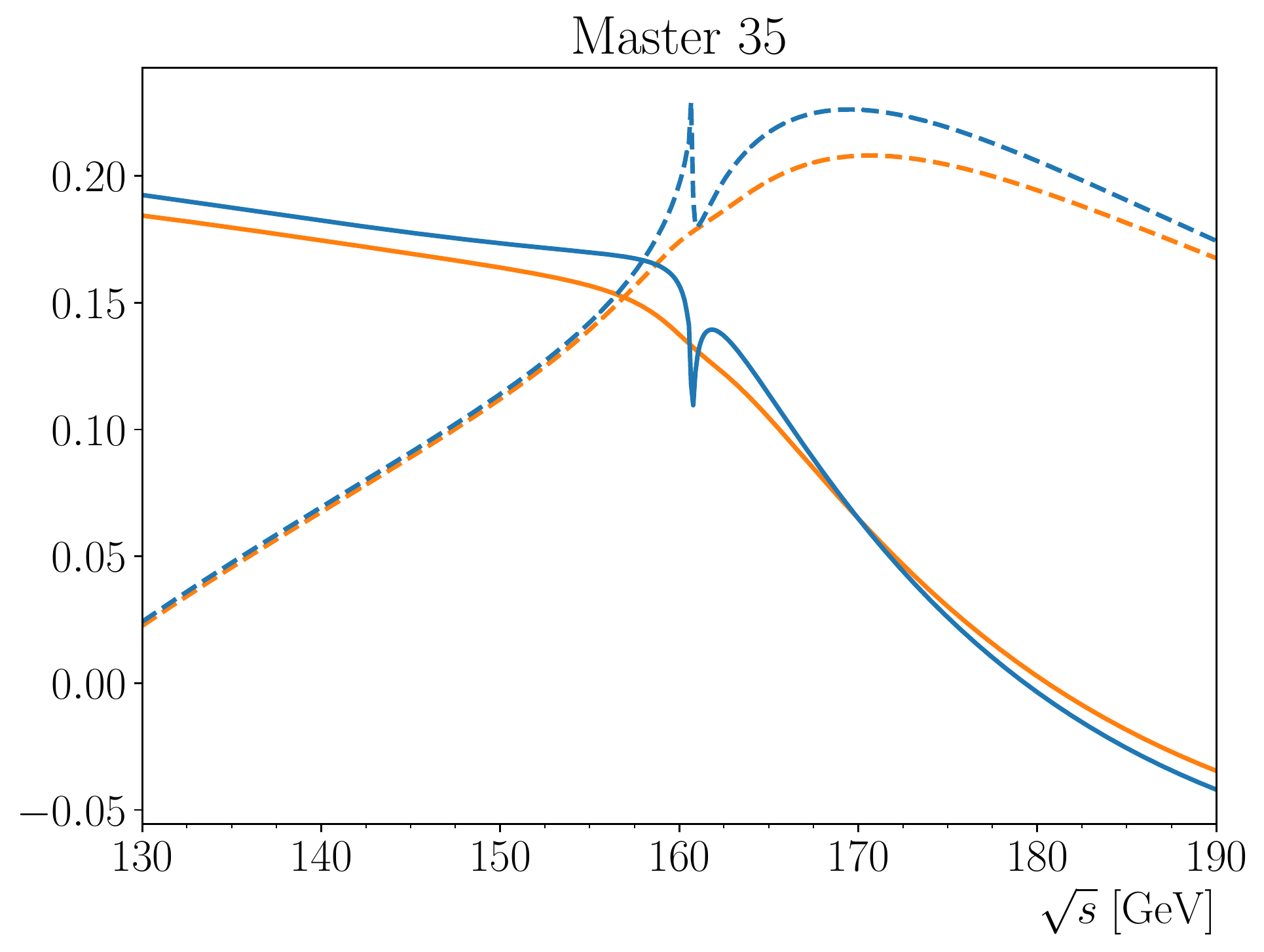}
    \includegraphics[width=0.45\textwidth]{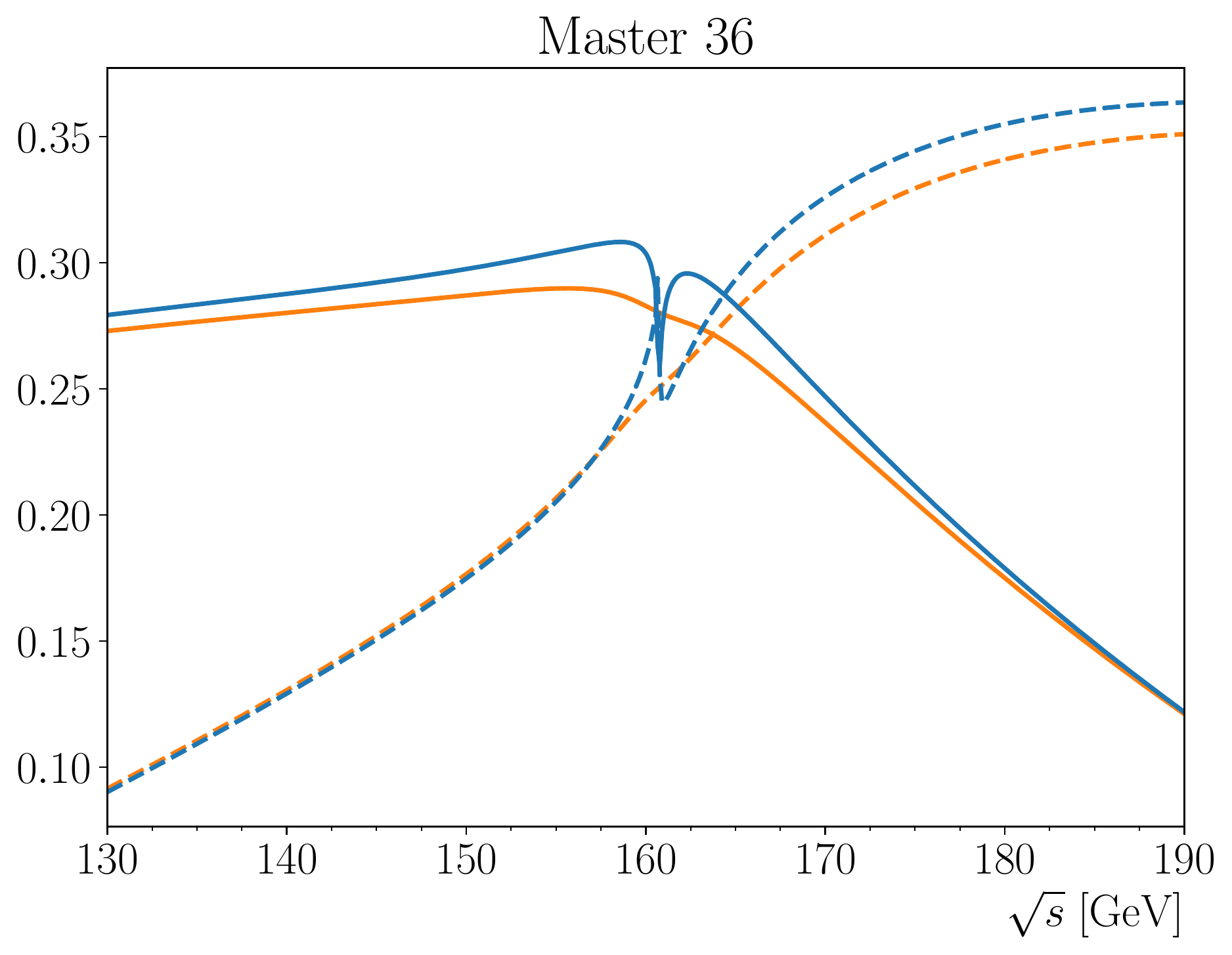}
    \caption{Comparison between real- and complex-valued masses for MI10 and MIs from 32 to 36. The plots show the value of the $\varepsilon^0$ order  for $130 \; \text{GeV} \le \sqrt{s} \le 190\;\text{GeV}$, $\cos\theta=0$ and $M^2=m_W^2$ or $M^2=\mu_W^2$. 
    In the MI10 plot, the red crosses represent the analytical value obtained with complex masses with \textsc{GiNaC}.}
    \label{fig:compReCoMass}
\end{figure}

In Fig.~\ref{fig:compReCoMass} we report the comparison between real- and complex-valued internal masses for six selected MIs, evaluated with {\sc SeaSyde}.
The plots present the MIs as a function of $\sqrt{s}$ at fixed $\cos\theta=0$, where we label with $\theta$ the scattering angle in the centre of mass reference frame.
The latter can be written in terms of the kinematic invariant $t$ introduced in Eq.~(\ref{eq:kinVariablesReal}) with the following relation:
\be
    t = -\frac{s}{2} \left(1-\cos\theta \right).
    \label{eq:relationTCostheta}
\ee
For each plot in Fig.~\ref{fig:compReCoMass}, the blue lines represent the masters for real-valued masses, while the orange the complex-valued ones. The solid and dashed lines represent the real and imaginary part of the solution, respectively.

The first plot shows the result for MI 10, defined as follows:
\be
    \text{Master 10}\;:\:\{1, 1, 2, 0, 0, 0, 1, 0, 0\}\;.
\ee
It provides a validation of our code: for the first 31 MIs, indeed, we have an analytic expression in terms of GPLs that can be checked against {\sc SeaSyde}, also for complex values of the internal mass. 
In the plot we can observe an excellent agreement between our result and the red crosses, that represent the value of the analytic result, evaluated with {\sc GiNaC} \cite{Bauer:2000cp,Vollinga:2004sn} for a few selected points.

The remaining 5 plots show the comparison between real- and complex-valued internal masses for MI32-36, and represent an original result of this paper.
We can observe that the masters with real-valued masses exhibit a non differentiable behaviour around the internal threshold $\sqrt{s}=2 m_W$, while they become smooth when moving to the CMS, making the latter particularly important for phenomenological studies in the threshold region.

\begin{figure}[t!]
    \centering
    \includegraphics[width=0.7 \textwidth]{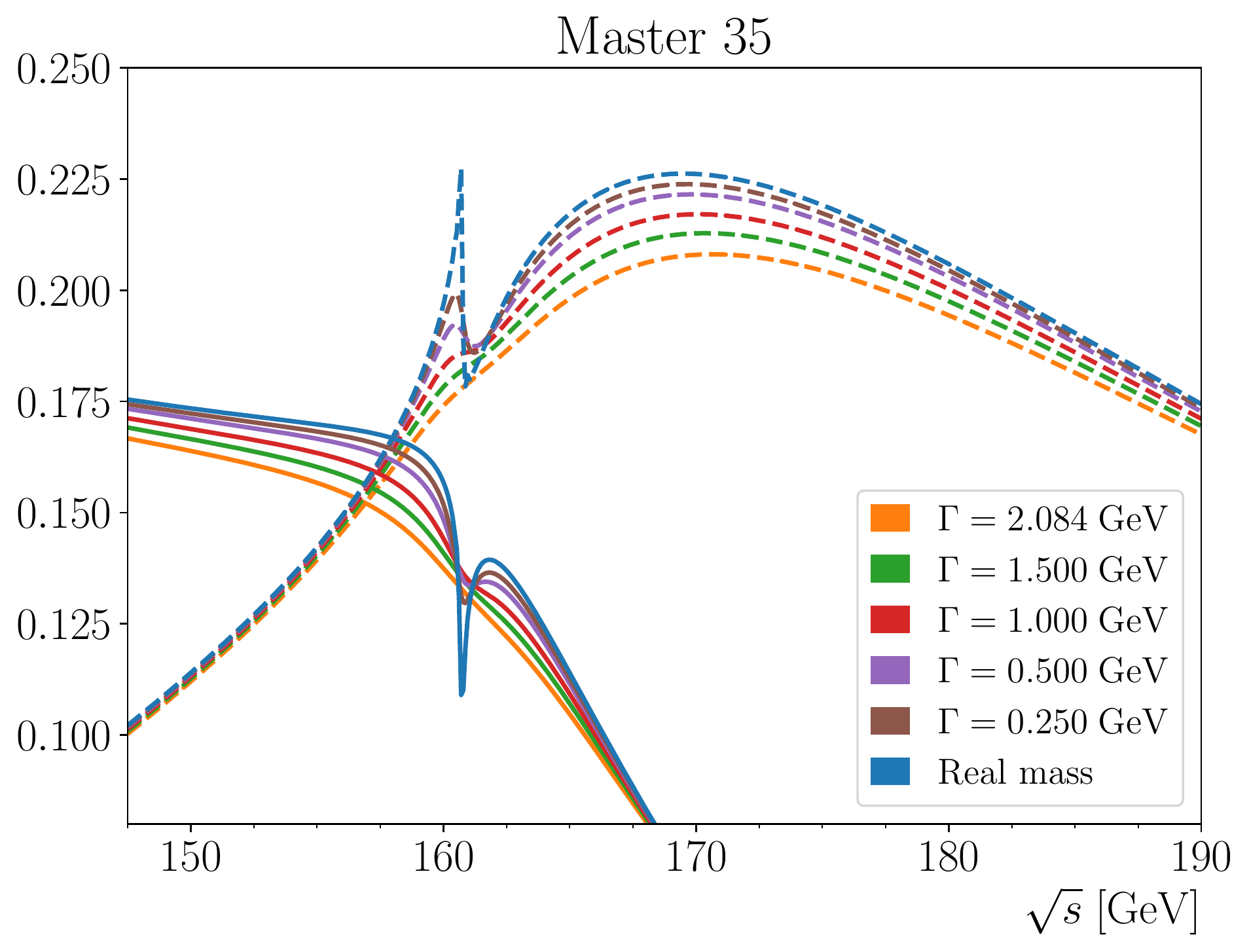}
    \caption{Comparison between different values of the decay rate $\Gamma$ for MI35. The plot shows the value of the $\varepsilon^0$ order for $150\;\text{GeV}\le\sqrt{s}\le190\;\text{GeV}$ and $\cos{\theta}=0$ with the mass of the $W$ boson. The blue line corresponds to the real-valued mass case, while the other colours are for the complex-valued case with different $\Gamma$ values. The solid and dashed lines represent the real and imaginary part of the solution.}
    \label{fig:limit}
\end{figure}

The definition of the complex mass given in Eq.~(\ref{eq:complex_mass}) shows that the case of real masses can be thought also as the limit for $\Gamma\to0$. Hence, by considering values of $\Gamma$ gradually smaller, all the MIs should converge to the one obtained with real-valued internal masses. This is shown explicitly for MI35 in Fig.~\ref{fig:limit}, although it holds for all the other MIs. In particular we have compared the results for real- and complex-valued masses, where the latter have been calculated using different values of the decay-width $\Gamma$. In the plot we observe that all the lines approach the blue one, representing the real mass case, as expected.
We note that in the case of complex internal masses the dimensionless kinematic variables $(x,y)$ become complex. Hence, a Feynman prescription is not necessary, since we can always link two points in the complex plane in a consistent way, as described in Section~\ref{subsec:path}. On the contrary, when working with real-valued masses, $(x,y)$ may lay on a branch-cut, namely the real axis. In this case providing a Feynman prescription is mandatory for performing the analytic continuation in an unambiguous way. In conclusion, this example besides showing that \textsc{SeaSyde} can robustly handle arbitrary internal complex masses, indicates also that the complex mass scheme is consistent with the Feynman prescription, and, moreover, it can be used to automatically implement the causality of the theory.

\subsection{Checks and performances}

As anticipated in the previous section, we have checked our numerical results against other publicly available tools, namely \textsc{GiNaC}, \textsc{DiffExp}, \textsc{AMFlow} \cite{Liu:2022chg}. In particular, we have used \textsc{DiffExp} and/or \textsc{AMFlow} to check the value of the 36 MIs considering internal real-valued masses in all the regions of the phase space. We also have compared our results against \textsc{pySecDec} \cite{Borowka:2017idc}, but only in few points of the Euclidean region, with a limited number of significant digits.
We have used \textsc{GiNaC} to check the first 31 masters with internal complex-valued masses and arbitrary values of the Mandelstam invariants. The value of the last 5 MIs with internal complex masses is, instead, a new prediction. 

The strength of the series expansion approach is that we can achieve an arbitrary level of precision, by simply adding more terms in the expansion. The increase in precision, however, corresponds to an increase in the amount of time and computational resources necessary to carry out the calculation. The two aspects, hence, have to be carefully balanced.

As a check, we evaluated the MIs in different points of the phase space, below and above the thresholds at $\sqrt{s}=m_V$ and $\sqrt{s}=2 m_V$ and for different values of $\cos\theta$, using a different number of terms in the series. In particular, using $150$ terms, we have obtained perfect agreement between \textsc{SeaSyde}, \textsc{DiffExp} and \textsc{GiNaC} up to the 40th decimal digit, i.e. quadruple floating-point precision. 

Regarding the time performances, they strongly depend not only on the number of terms in the series, but also on the presence of singularities between the starting and ending points. In Table \ref{table:checks} we report the time necessary for transporting the boundary conditions from the Euclidean region to a test point, namely $\sqrt{s}=155$ GeV and $\cos\theta=0$, for different number of terms, along with the precision of the result.

\begin{table}[ht!]
    \centering
    \begin{tabular}{c c c}
    \toprule
    Number of terms & Precision & Execution time\\
    \toprule
    50 terms & $10^{-14}$ & $\sim 14$ min\\
    \midrule
    75 terms & $10^{-19}$ & $\sim 26$ min\\
    \midrule
    100 terms & $10^{-25}$ & $\sim 50$ min\\
    \midrule
    125 terms & $10^{-33}$ & $\sim 75$ min\\
    \midrule
    150 terms & $10^{-40}$ & $\sim 90$ min\\
    \toprule
    \end{tabular}
    \caption{Precision achieved and execution time for transporting the boundary conditions from the Euclidean region to $\sqrt{s}=155$ GeV and $\cos\theta=0$, for different number of terms in the series.}
    \label{table:checks}
\end{table}

This estimate is obtained on a MacBook Pro (2015) equipped with a 2.7 GHz Dual-Core Intel i5. The fact that the execution time increases so much with the number of terms, is mainly due to the poor-performances of the \textsc{Mathematica} function \texttt{Series} when dealing with series with a big number of terms. Anyhow, for most current precision studies, a $10^{-19}$ precision is sufficient. The execution time depends also on the distance between the starting point, where the boundary conditions are imposed, and the final one, in which we would like to evaluate the solution. For reaching the physical region we usually need a number between $15$ and $20$ steps. Given the results in Table \ref{table:checks} we can estimate a total of $50$, $160$ and $300$ seconds per step for $50$, $100$ and $150$ terms, respectively.
Once in the physical region, less singularities need to be avoided and, hence, more direct path can be chosen. In this case a number between $1$ and $10$ steps is sufficient to reach every point in the phase-space.

Given the time performances of our package, and in general of packages implementing the series expansion approach, a direct implementation in Monte Carlo generators is out of question.
However it is possible to reformulate the problem in such a way that an evaluation on-the-fly is not necessary: we can use {\sc SeaSyde} to obtain a numerical grid for the MIs first; a grid for the total correction can be computed later, covering the whole phase space of the process;
once this latter grid is ready, we can obtain the value of the correction at any point, simply by a linear interpolation, thanks to the smoothness of the total correction.
In Ref.~\cite{Armadillo:2022bgm} we presented a numerical grid for the MI32-36. This grid is constituted by $(130\times25)$ points which samples the phase-space for different values of $(\sqrt{s},\cos\theta)$ in the range $\sqrt{s}\in [40,13000 ] $ GeV and $\cos\theta\in[-1,1]$. The full evaluation was performed in parallel on 26 cores on an Intel Xeon Silver 4110 2.1 GHz, considering $75$ terms, and took 11 hours in total. 
  The development of interpolation techniques is an active field of research
  (see e.g. \cite{Chawdhry:2019bji,Bishara:2019iwh,Winterhalder:2021ngy} ),
  in part because of the increasing evaluation time per phase-space point of the two-loop virtual amplitudes.
  A detailed discussion of these alternatives is however beyond the scope of this paper.

\section{Conclusions}
\label{sec:conclusion}
The solution via series expansion of multi-loop Feynman integrals has received a renewed attention after the work of Ref.~\cite{Moriello:2019yhu} and the publication of the package \textsc{DiffExp} \cite{Hidding:2020ytt}.
The definition of the masses of the gauge bosons in the Complex-Mass-Scheme implies the need to evaluate multi-loop Feynman integrals with complex-valued internal massive lines and the latter could not be studied with the public version of \textsc{DiffExp}.
In this paper we have presented a new implementation of the series expansion method in the package \textsc{SeaSyde}, with an original algorithm to perform the analytic continuation of the results, in the whole complex plane of the kinematical invariants.
The usage of this package allows the consistent evaluation of two-loop and higher-order electroweak corrections to scattering processes of interest at the LHC and future colliders.

\section*{Acknowledgments}
S.D. and A.V. are supported by the Italian Ministero della Universit\`a e della Ricerca (grant PRIN201719AVICI\_01).
R. B. is partly supported by the italian Ministero della Universit\`a e della Ricerca (MIUR) under grant PRIN 20172LNEEZ.
T.A. acknowledges the support of the F.R.S.-FNRS under the grant F.R.S.-FNRS–PDR–T.0142.18.

\appendix

\section{Package documentation}
\label{app:packagedoc}

The \textsc{SeaSyde} package (Series Expansion Approach for SYstems of Differential Equations) can be imported in Mathematica using the command \texttt{Get[\dots]}, i.e. through \texttt{<< SeaSyde.m}. Note that it has been developed and tested on \textsc{Mathematica 12.0}, no previous version of \textsc{Mathematica} has been tested. 

Below we present all different functions and their functionalities. 
\begin{itemize}
    \item \texttt{CurrentConfiguration[]}
    
    It returns the current configuration of the package. The configuration parameters can be modified using the function \texttt{UpdateConfiguration}.
    
    \item \texttt{UpdateConfiguration[NewConfig\_]}
    
    \texttt{NewConfig} must be a \texttt{List} whose elements are replacement rules \texttt{NameParameter -> NewValue}. See Table~\ref{table:configuration} for a complete overview on all the parameters that the user can modify.
    
   \begin{table}[t!]
    \centering
      \makebox[\linewidth]{
      \centering
       \begin{tabular}{|C{4 cm}|C{2 cm}| C{1.5 cm}|J{7 cm}|}
       \hline
        NameParameter & Value type & Default & Description \\
        \hline
        {\small\texttt{EpsilonOrder}} & \texttt{Integer} & 2 & {\small The maximum order in the dimensional regulator $\varepsilon$ at which the system is expanded. Note that the minimum order is determined by the boundary conditions, i.e. if they contain a term $1/\varepsilon^2$ the minimum order will be $-2$.} \\
        \hline
       {\small \texttt{ExpansionOrder}} & \texttt{Integer} & 50 & {\small The maximum order in the kinematic variable at which the solution is expanded.}\\
        \hline
        {\small\texttt{InternalWorkingPrecision}} & \texttt{Integer} & 250 & {\small Specifies the number of digits that are used in internal calculations. If it is too high, the execution will require more time and space, if it is too low, we may face rounding errors.} \\
        \hline
        {\small\texttt{ChopPrecision}} & \texttt{Integer} & 150 & {\small Replaces number which are smaller than $10^{-\texttt{ChopPrecision}}$ by exactly $0$ in internal computations.} \\
        \hline
        {\small\texttt{LogarithmicExpansion}} & \texttt{Bool} & \texttt{False} & {\small It specifies which method to use for transporting boundary conditions. If it is set to \texttt{True}, \textsc{SeaSyde} will expand also on top of singularities.}\\
        \hline
        {\small\texttt{RadiusOfConvergence}} & \texttt{Integer} & 2 & {\small It controls how fast we move at every expansion. If \texttt{RadiusOfConvergence} is $n$, and the maximum radius of convergence of the solution is $r$, then the new point will be distant $r/n$ from the center of the series. Note that $r$ is determined internally at every step, based on the position of singularities.}\\
        \hline
        {\small\texttt{LogarithmicSingularities}} & \texttt{List} & \texttt{\{\}} & {\small The user can explicitly state which singularities are of Logarithmic type, i.e. develop a branch-cut, and which ones are not. This might speed up the evaluation of numerical grids since it allows more direct paths in the phase-space. The format in which the singular points are passed must the one returned by \texttt{CheckSingularities[]}.}\\
        \hline
        {\small\texttt{SafeSingularities}} & \texttt{List} & \texttt{\{\}} & {\small Same as \texttt{LogarithmicSingularities}.}\\
        \hline
       \end{tabular}
       }
       \caption{All the parameters that can be modified by the user.}
       \label{table:configuration}
   \end{table}
    
    \item \texttt{ReadFrom[FilePath\_]} 
    
    Utility function which reads from the specified path and returns the content of the file.
    
    \item \texttt{SetSystemOfDifferentialEquation[System\_, BCs\_, MIs\_, Variables\_, PointBC\_, Param\_:\{\}]}
    
    It sets all the internal variables of the package and prepare the system of differential equations. It receives in input
    \begin{itemize}
        \item \texttt{System\_}: the system of differential equations. The system equations must be given in triangular form and it must be ordered so that, order by order in $\varepsilon$, every equation contains only MIs from previous equations. If there are multiple kinematic variables, for example \texttt{x} and \texttt{y}, the first $n$ equations in \texttt{System} must be the ones with respect to \texttt{x}, while the last $n$ the ones with respect to \texttt{y}, where $n$ is the number of MIs. The equations can be given expanded in $\varepsilon$ or in a closed form in $\varepsilon$.\\
        \textit{Example:}
\begin{Verbatim}[commandchars=\\\{\}]
\{  B\textsubscript{1}\textsuperscript{(1,0)}[x,y] == 0 ,
    B\textsubscript{2}\textsuperscript{(1,0)}[x,y] == (- 1/x - \textepsilon/x)B\textsubscript{2}[x,y] - \textepsilon B\textsubscript{1}[x,y] / x,
    B\textsubscript{1}\textsuperscript{(0,1)}[x,y] == 0 ,
    B\textsubscript{2}\textsuperscript{(0,1)}[x,y] == 0 \}
\end{Verbatim}
\item \texttt{BCs\_}: the boundary conditions for the given equation. They can be given in a closed form in $\varepsilon$ or as a series. They can also be given as an asymptotic limit. They can be exact or floating numbers, in this case make sure that the precision of the boundary condition is sufficient for your final precision goal. Note that the argument of the function is ignored, hence, the point in which the boundary conditions are imposed is only fixed by the \texttt{PointBC} parameter.
        \\
        \textit{Example:}
\begin{Verbatim}[commandchars=\\\{\}]
\{  B\textsubscript{1}[1,1] == - 1/32 + \textepsilon/32 + \textepsilon\textsuperscript{2}( - 1/32 + \textpi\textsuperscript{2}/96 ),
    B\textsubscript{2}[1,1] == - \textepsilon Log[2]/16 + \textepsilon\textsuperscript{2}(-\textpi\textsuperscript{2}/192 + Log[2]\textsuperscript{2}/8) \}
\end{Verbatim}
        
        \item \texttt{MIs\_}: the list of master integrals, as they appear in the equations.\\
        \textit{Example:}
\begin{Verbatim}[commandchars=\\\{\}]
\{  B\textsubscript{1}[x,y], B\textsubscript{2}[x,y] \} 
\end{Verbatim}
        
        \item \texttt{Variables\_}: the list of variables appear in the equations, together with their Feynman prescriptions.\\
        \textit{Example:}
\begin{Verbatim}[commandchars=\\\{\}]
\{  x - I \textdelta, y + I \textdelta \} 
\end{Verbatim}       
        
        \item \texttt{PointBC\_}: the point in the phase-space in which the boundary conditions are imposed.\\
        \textit{Example:}
\begin{Verbatim}[commandchars=\\\{\}]
\{ 1, 1 \} 
\end{Verbatim}           
        
        \item \texttt{Param\_}: this is an optional parameter. Some equations might contain some external parameters, for example some masses \texttt{Mw}, \texttt{Mz}. This substitutions are performed before solving the system.\\
        \textit{Example:}
\begin{Verbatim}[commandchars=\\\{\}]
\{ MW -> 80.38, MZ -> 91.19 \} 
\end{Verbatim}        
    \end{itemize}
    
    \item \texttt{GetSystemOfDifferentialEquation[]} and\\ \texttt{GetSystemOfDifferentialEquationExpanded[]}
    
    They return the system of differential equations before and after it has been expanded in $\varepsilon$. These functions can be used to check if everything has been set correctly.
    
    \item \texttt{SolveSystem[Variable\_]} 
    
    It solves the system of differential equations with respect to the kinematic variable \texttt{Variable}. The series solution in centred in the point where the boundary conditions are imposed. After solving the system of differential equations, it is possible to obtain the solution through \texttt{Solution[]}, \texttt{SolutionValue[]} or \texttt{SolutionTable[]}.
    
    \item \texttt{GetPoint[]}
    
    It returns the current point in which the boundary conditions are imposed.
    
    \item \texttt{TransportVariable[Variable\_, Destination\_, Line\_:\{\}]}
    
    It transports the boundary conditions for the variable \texttt{Variable} from the current point to \texttt{Destination}. After transporting the boundary conditions, the point in which the boundary conditions are imposed is updated to \texttt{Destination}. The \texttt{Line} parameter is optional. If the user is not satisfied by the path automatically chosen by the package can use their own. The \texttt{Line} object must be created with \texttt{CreateLine}
    
    \item \texttt{CreateLine[Points\_]}
    
    It returns a line object that can be used in \texttt{TransportVariable}
    
    \item \texttt{TransportBoundaryConditions[PhaseSpacePoint\_]}
    
    \texttt{PhaseSpacePoint} must be a \texttt{List} whose length is given by the number of kinematic variables. Its first element must be the final value for the first variable, its second element the value for the second variable, and so on. The order of the kinematic variables is the same passed as an input in\texttt{SetSystemOfDifferentialEquation}. After transporting the boundary conditions, the point in which they are imposed is updated to \texttt{PhaseSpacePoint}.
    
    \item \texttt{Solution[]}
    
    It returns the series solution in the current point. The coefficients of the series are given with \texttt{InternalWorkingPrecision} digits. The result is given as a \texttt{List} and every MI as a Laurent series in $\varepsilon$.
    
    \item \texttt{SolutionValue[]}
    
    It returns the value of the MIs, in the centre of the series, as a Laurent expansion in $\varepsilon$. The coefficients of the $\varepsilon$-series are given with \texttt{InternalWorkingPrecision} digits.
    
    \item \texttt{SolutionTable[]}
    
    It returns the value of the MIs, in the centre of the series, as a \texttt{List} of \texttt{List} going from the first to the last master and from the minimum to the maximum order in $\varepsilon$. The coefficients of the $\varepsilon$-series are given with \texttt{InternalWorkingPrecision} digits.
    
    \item \texttt{CheckSingularities[]}
    
    It checks whether the singularities are logarithmic, i.e. if they develop a branch-cut. The check is done by performing a path round the singularity and checking if the solution has developed, or not, an imaginary part for at least one of the coefficients. If it is not, it means that in doing so we crossed a branch-cut and, hence, the singularity is logarithmic. If this function is not called, \texttt{SeaSyde} consider every singularity as logarithmic, and the analytic continuation is still possible. However, if the user is planning to make an intensive use of the function \texttt{TransportBoundaryConditions}, e.g. create a numerical grid, knowing the position of non-logarithmic singularities might allow for more direct paths in the complex plane. The output of \texttt{CheckSingularities} can be passed back in the \texttt{SafeSingularities} and \texttt{LogarithmicSingularities} parameters for future runs.

    
    
    
    
    \item {\texttt{CreateGraph[MI\_, EpsOrder\_, Left\_, Right\_, OtherFunctions\_:\{\}]}}
    
    Draws a \texttt{ReImPlot} of the solution of order \texttt{EpsOrder} for the master \texttt{MI}. The graph runs from \texttt{Left} to \texttt{Right}. 
    The argument \texttt{OtherFunctions} may contains other functions to be plotted in the same graph.
\end{itemize}

In the \texttt{Example/} folder of the GitHub repository of \texttt{SeaSyde}, the user can find working examples to play with.

\section{Exact boundary conditions}
\label{app:exactbcs}
We report the exact boundary conditions we used for solving the 36x36 system of differential equations for the neutral-current Drell-Yan. The MIs are labelled $\mi_i$ and numbered as in Ref.~\cite{Bonciani:2016ypc}. The boundary conditions are taken in the asymptotic limit 
\be
x\to 0^+\;, \qquad y=1,
\ee
where $x$ and $y$ are the kinematic variables introduced in Section \ref{sec:results}. 
\vspace*{3mm}

\vspace*{3mm}

\begingroup
\allowdisplaybreaks
\begin{align}
%
\mi_1 &=   \frac{1}{16 \ep^2}  \bigg[
-\frac{1}{2}
-\frac{\ep}{2}
-\ep^2 \Big(
         \frac{1}{2}
        +\zeta_2
\Big)
+\ep^3 \Big(
        -\frac{1}{2}
        -\zeta_2
        +\zeta_3
\Big)
\nonumber\\& \qquad \quad
+\ep^4 \Big(
        -\frac{1}{2}
        -\zeta_2
        -\frac{9}{5} \zeta_2^2
        +\zeta_3
\Big)
\bigg],
\nonumber\\
\mi_2 &=   \frac{1}{16 \ep^2}  \bigg[
\ep
+\ep^2
+\ep^3 \Big(
        1
        +2 \zeta_2
\Big)
+\ep^4 \Big(
        1
        +2 \zeta_2
        -2 \zeta_3
\Big)
\bigg],
\nonumber\\
\mi_3 &=   \frac{1}{16 \ep^2}  \bigg[
-1
-2 \ep
+\ep^2 \Big(
        -2
        -2 \zeta_2
\Big)
+\ep^3 \Big(
        -2
        -4 \zeta_2
        +2 \zeta_3
\Big)
\nonumber\\& \qquad \quad
+\ep^4 \Big(
        -2
        -4 \zeta_2
        -\frac{18}{5} \zeta_2^2
        +4 \zeta_3
\Big)
\bigg],
\nonumber\\
\mi_4 &=   \frac{1}{16 x \ep^2}  \bigg[
1
-\ep \log (x)
+\ep^2 \Big(
        \frac{\log ^2(x)}{2}
        -\zeta_2
\Big)
+\ep^3 \Big(
        -2 \zeta_3
        +\zeta_2 \log (x)
        -\frac{1}{6} \log ^3(x)
\Big)
\nonumber\\& \qquad \quad
+\ep^4 \Big(
        -\frac{9}{10} \zeta_2^2
        +2 \zeta_3 \log (x)
        -\frac{1}{2} \zeta_2 \log ^2(x)
        +\frac{\log ^4(x)}{24}
\Big)
\bigg],
\nonumber\\
\mi_5 &=   \frac{1}{16 \ep^2}  \bigg[
-1
+2 \ep^2 \zeta_2
+10 \ep^3 \zeta_3
+\frac{22}{5} \ep^4 \zeta_2^2
\bigg],
\nonumber\\
\mi_6 &=   \frac{1}{16 x \ep^3}  \bigg[
\frac{1}{4}
-\frac{1}{2} \ep \log (x)
+\frac{1}{2} \ep^2 \log ^2(x)
+\ep^3 \Big(
        -\frac{1}{3} \log ^3(x)
        -2 \zeta_3
\Big)
\nonumber\\& \qquad \quad
+\ep^4 \Big(
        -\frac{6}{5} \zeta_2^2
        +4 \zeta_3 \log (x)
        +\frac{\log ^4(x)}{6}
\Big)
\bigg],
\nonumber\\
\mi_7 &=   \frac{1}{16 \ep^3}  \bigg[
\frac{1}{4}
-2 \ep^3 \zeta_3
-\frac{6}{5} \ep^4 \zeta_2^2
\bigg],
\nonumber\\
\mi_8 &=   \frac{1}{16 \ep^3}  \bigg[
\ep^2
+\ep^3
+\ep^4 \Big(
        1
        +2 \zeta_2
\Big)
\bigg],
\nonumber\\
\mi_9 &=   \frac{1}{16 \ep^2}  \bigg[
-\frac{\ep}{2}
-\frac{3 \ep^2}{2}
+\ep^3 \Big(
        -\frac{3}{2}
        -\zeta_2
\Big)
+\ep^4 \Big(
        -\frac{3}{2}
        -3 \zeta_2
        +\zeta_3
\Big)
\bigg],
\nonumber\\
\mi_{10} &=   \frac{1}{16 \ep^2}  \bigg[
\ep
+\ep^2
+\ep^3 \Big(
        1
        +2 \zeta_2
\Big)
+\ep^4 \Big(
        1
        +2 \zeta_2
        -2 \zeta_3
\Big)
\bigg],
\nonumber\\
\mi_{11} &=   \frac{1}{16 x \ep^2}  \bigg[
-\frac{\ep}{2}
+\frac{1}{2} \ep^2 \log (x)
+\ep^3 \Big(
        -\frac{1}{4} \log ^2(x)
        +\frac{\zeta_2}{2}
\Big)
\nonumber\\& \qquad \quad
+\ep^4 \Big(
        -\frac{1}{2} \zeta_2 \log (x)
        +\frac{\log ^3(x)}{12}
        +\zeta_3
\Big)
\bigg],
\nonumber\\
\mi_{12} &=   \frac{1}{16 \ep^3}  \bigg[
\ep^2 \zeta_2
+2 \ep^3 \zeta_3
+2 \ep^4 \zeta_2^2
\bigg],
\nonumber\\
\mi_{13} &=   \frac{1}{16 \ep^4}  \bigg[
\ep^4 \Big(
        -3
        -2 \zeta_2
        +2 \log (x)
        -\frac{1}{2} \log ^2(x)
\Big)
\bigg],
\nonumber\\
\mi_{14} &=   \frac{1}{16 x \ep^3}  \bigg[
\frac{\ep}{2}
+\ep^3 \Big(
        -1
        -\zeta_2
        +\log (x)
        -\frac{1}{2} \log ^2(x)
\Big)
\nonumber\\& \qquad \quad
+\ep^4 \Big(
        -3
        -\zeta_2
        +2 \zeta_3
        +\Big(
                3
                +\zeta_2
        \Big) \log (x)
        -\frac{3}{2} \log ^2(x)
        +\frac{\log ^3(x)}{2}
\Big)
\bigg],
\nonumber\\
\mi_{15} &=   \frac{1}{16 \ep^4}  \bigg[
\ep^3
+\ep^4
\bigg],
\nonumber\\
\mi_{16} &=   \frac{1}{16 \ep^3}  \bigg[
\frac{\ep}{2}
+\frac{3 \ep^2}{2}
+\ep^4 \Big(
        \frac{3}{2}
        +3 \zeta_2
        -\zeta_3
\Big)
+\ep^3 \Big(
        \frac{3}{2}
        +\zeta_2
\Big)
\bigg],
\nonumber\\
\mi_{17} &=   \frac{1}{16 \ep^3}  \bigg[
\ep^3 (-2+\log (x))
+\ep^4 \Big(
        -6
        +3 \zeta_2
        +3 \log (x)
        -\frac{1}{2} \log ^2(x)
\Big)
\bigg],
\nonumber\\
\mi_{18} &=   \frac{1}{16 x \ep^3}  \bigg[
-\ep^2
+\ep^4 \Big(
        -1
        +\log (x)
        -\frac{1}{2} \log ^2(x)
        +\zeta_2
\Big)
+\ep^3 (-1+\log (x))
\bigg],
\nonumber\\
\mi_{19} &=   \frac{1}{16 x \ep^3}  \bigg[
-\frac{9}{4}
+\ep^3 \Big(
        24 \zeta_3
        -12 \zeta_2 \log (x)
        -\frac{1}{2} \log ^3(x)
\Big)
\nonumber\\& \qquad \quad
+\ep^4 \Big(
        -18 \zeta_3 \log (x)
        +6 \zeta_2 \log ^2(x)
        +\frac{3 \log ^4(x)}{8}
\Big)
+12 \ep^2 \zeta_2
+\frac{3}{2} \ep \log (x)
\bigg],
\nonumber\\
\mi_{20} &=   \frac{1}{16 \ep^4}  \bigg[
2 \ep^3 \zeta_3
+\frac{13}{5} \ep^4 \zeta_2^2
\bigg],
\nonumber\\
\mi_{21} &=   \frac{1}{16 \ep^3}  \bigg[
-\frac{1}{4}
+\ep^2 \zeta_2
+5 \ep^3 \zeta_3
+\frac{27}{5} \ep^4 \zeta_2^2
\bigg],
\nonumber\\
\mi_{22} &=   \frac{1}{16 \ep^4}  \bigg[
2 \ep^3 \zeta_3
+\frac{13}{5} \ep^4 \zeta_2^2
\bigg],
\nonumber\\
\mi_{23} &=   \frac{1}{16 \ep^3}  \bigg[
-\frac{1}{2}
+3 \ep^2 \zeta_2
+13 \ep^3 \zeta_3
+\frac{57}{5} \ep^4 \zeta_2^2
\bigg],
\nonumber\\
\mi_{24} &=   \frac{1}{16 \ep^3}  \bigg[
-\ep^2 \zeta_2
+\ep^3 \Big(
        -2 \zeta_2
        -3 \zeta_3
\Big)
+\ep^4 \Big(
        -4 \zeta_2
        -\frac{21}{5} \zeta_2^2
        -6 \zeta_3
\Big)
\bigg],
\nonumber\\
\mi_{25} &=   \frac{1}{16 \ep^3}  \bigg[
-\frac{1}{4}
+\ep^2 \zeta_2
+5 \ep^3 \zeta_3
+\frac{27}{5} \ep^4 \zeta_2^2
\bigg],
\nonumber\\
\mi_{26} &=   \frac{1}{16 \ep^3}  \bigg[
-\ep^2
+\ep^3 \Big(
        -4
        +\zeta_2
\Big)
+\ep^4 \Big(
        -12
        +2 \zeta_2
        +2 \zeta_3
\Big)
\bigg],
\nonumber\\
\mi_{27} &=   \frac{1}{16 \ep^3}  \bigg[
-\ep^2
+\ep^3 \Big(
        2
        -3 \zeta_2
\Big)
+\ep^4 \Big(
        -4
        +4 \zeta_2
        -6 \zeta_3
\Big)
\bigg],
\nonumber\\
\mi_{28} &=   \frac{1}{16 \ep^3}  \bigg[
\frac{1}{4}
+\ep^2 \Big(
        -1
        -\zeta_2
\Big)
+\ep^3 \Big(
        1
        -3 \zeta_2
        -5 \zeta_3
\Big)
+\ep^4 \Big(
        -1
        -\frac{27}{5} \zeta_2^2
        -9 \zeta_3
        +\zeta_2
\Big)
\bigg],
\nonumber\\
\mi_{29} &=   \frac{1}{16 \ep^4}  \bigg[
\ep^4 \Big(
        1
        +2 \zeta_2
        -\log (x)
        +\frac{\log ^2(x)}{2}
\Big)
\bigg],
\nonumber\\
\mi_{30} &=   \frac{1}{16 x \ep^4}  \bigg[
 \ep^2 \zeta_2
-\ep^3 \zeta_2 \log (x)
+\ep^4 \Big(
        -\frac{43}{10} \zeta_2^2
        +\frac{1}{2} \zeta_2 \log ^2(x)
\Big)
\bigg],
\nonumber\\
\mi_{31} &=   \frac{1}{16 x \ep^3}  \bigg[
-\ep
+\ep^2 \Big(
        3
        +\log (x)
        -2 \zeta_2
\Big)
+\ep^3 \Big(
        -9
        +11 \zeta_2
        -\Big(
                 3
                -2 \zeta_2
        \Big) \log (x)
        -\frac{1}{2} \log ^2(x)
\Big)
\nonumber\\& 
+\ep^4 \Big(
        27
        -33 \zeta_2
        +\frac{43}{5} \zeta_2^2
        +26 \zeta_3
        +\Big(
                9
                -11 \zeta_2
        \Big) \log (x)
        +\Big(
                \frac{3}{2}
                -\zeta_2
        \Big) \log ^2(x)
        +\frac{\log ^3(x)}{6}
\Big)
\bigg],
\nonumber\\
\mi_{32} &=   \frac{1}{16 \ep^4}  \bigg[
-\ep^3 \zeta_2
-6 \ep^4 \zeta_3
\bigg],
\nonumber\\
\mi_{33} &=   \frac{1}{16 x \ep^4}  \bigg[
-\ep^2
+\ep^3 \Big(
        3
        +\log (x)
        -2 \zeta_2
\Big)
+\ep^4 \Big(
        -9
        +11 \zeta_2
        -\Big(
                 3
                -2 \zeta_2
        \Big) \log (x)
\nonumber\\& \qquad \quad        
        -\frac{1}{2} \log ^2(x)
\Big)
\bigg],
\nonumber\\
\mi_{34} &=   \frac{1}{16 x \ep^4}  \bigg[
\ep^2
+\ep^3 \Big(
        3
        -\log (x)
        -2 \zeta_2
\Big)
+\ep^4 \Big(
        5
        -7 \zeta_2
        +\Big(
                -3
                +2 \zeta_2
        \Big) \log (x)
        +\frac{\log ^2(x)}{2}
\Big)
\bigg],
\nonumber\\
\mi_{35} &=   \frac{1}{16 x \ep^4}  \bigg[
\ep^2
+\ep^3 \Big(
        -3
        -\log (x)
        +4 \zeta_2
\Big)
+\ep^4 \Big(
        9
        -11 \zeta_2
        +12 \zeta_3
        +\Big(
                3
                -4 \zeta_2
        \Big) \log (x)
\nonumber\\& \qquad \quad        
        +\frac{\log ^2(x)}{2}
\Big)
\bigg],
\nonumber\\
\mi_{36} &=   \frac{1}{16 x \ep^4}  \bigg[
-\ep^2
+\ep^3 \Big(
        -1
        +\log (x)
        +2 \zeta_2
        -4 \zeta_3
\Big)
+\ep^4 \Big(
        -1
        -\frac{33}{5} \zeta_2^2
        +12 \zeta_3
\nonumber\\& \qquad \quad        
        +\Big(
                1
                -2 \zeta_2
                +4 \zeta_3
        \Big) \log (x)
        -\frac{1}{2} \log ^2(x)
        +\zeta_2
\Big)
\bigg]\nonumber,
\end{align}
\endgroup
where $\zeta_n=\zeta(n)$ denotes the \textit{Riemann zeta function}.

\bibliographystyle{elsarticle-num}
\bibliography{SeaSyde}

\end{document}